\pdfoutput=1

\documentclass[11pt,twoside,a4paper,cmspaper,final,collab]{cms-tdr}
\def\svnVersion{87e1dfa}\def\svnDate{2020/09/07}\def\cmsCernNoTag{CERN-EP-2019-260}\def\cmsCernDate{\today}\def\cmsMessage{'Published in the European Physical Journal C as \href{http://dx.doi.org/10.1140/epjc/s10052-020-08562-y}{\doi{10.1140/epjc/s10052-020-08562-y}.}'}

\begin{document}\cmsNoteHeader{FSQ-12-033}

\hyphenation{had-ron-i-za-tion}
\hyphenation{cal-or-i-me-ter}
\hyphenation{de-vices}
\RCS$HeadURL$
\RCS$Id$
\newlength\cmsFigWidth
\newlength\cmsFigWidthSingleSmall
\newlength\cmsFigWidthSingleLarge
\newlength\cmsFigWidthSmall
\ifthenelse{\boolean{cms@external}}{\setlength\cmsFigWidth{0.85\columnwidth}}{\setlength\cmsFigWidth{0.49\textwidth}}
\ifthenelse{\boolean{cms@external}}{\setlength\cmsFigWidthSingleSmall{0.85\columnwidth}}{\setlength\cmsFigWidthSingleSmall{0.45\textwidth}}
\ifthenelse{\boolean{cms@external}}{\setlength\cmsFigWidthSingleLarge{0.85\columnwidth}}{\setlength\cmsFigWidthSingleLarge{0.60\textwidth}}
\ifthenelse{\boolean{cms@external}}{\setlength\cmsFigWidthSmall{0.60\columnwidth}}{\setlength\cmsFigWidthSmall{0.40\textwidth}}
\ifthenelse{\boolean{cms@external}}{\providecommand{\cmsLeft}{upper\xspace}}{\providecommand{\cmsLeft}{left\xspace}}
\ifthenelse{\boolean{cms@external}}{\providecommand{\cmsRight}{lower\xspace}}{\providecommand{\cmsRight}{right\xspace}}

\cmsNoteHeader{FSQ-12-033}
\renewcommand{\cmsCollabName}{The CMS and TOTEM Collaborations}
\renewcommand{\cmsNUMBER}{FSQ-12-033}
\renewcommand{\cmslogo}{\includegraphics[height=2.33cm]{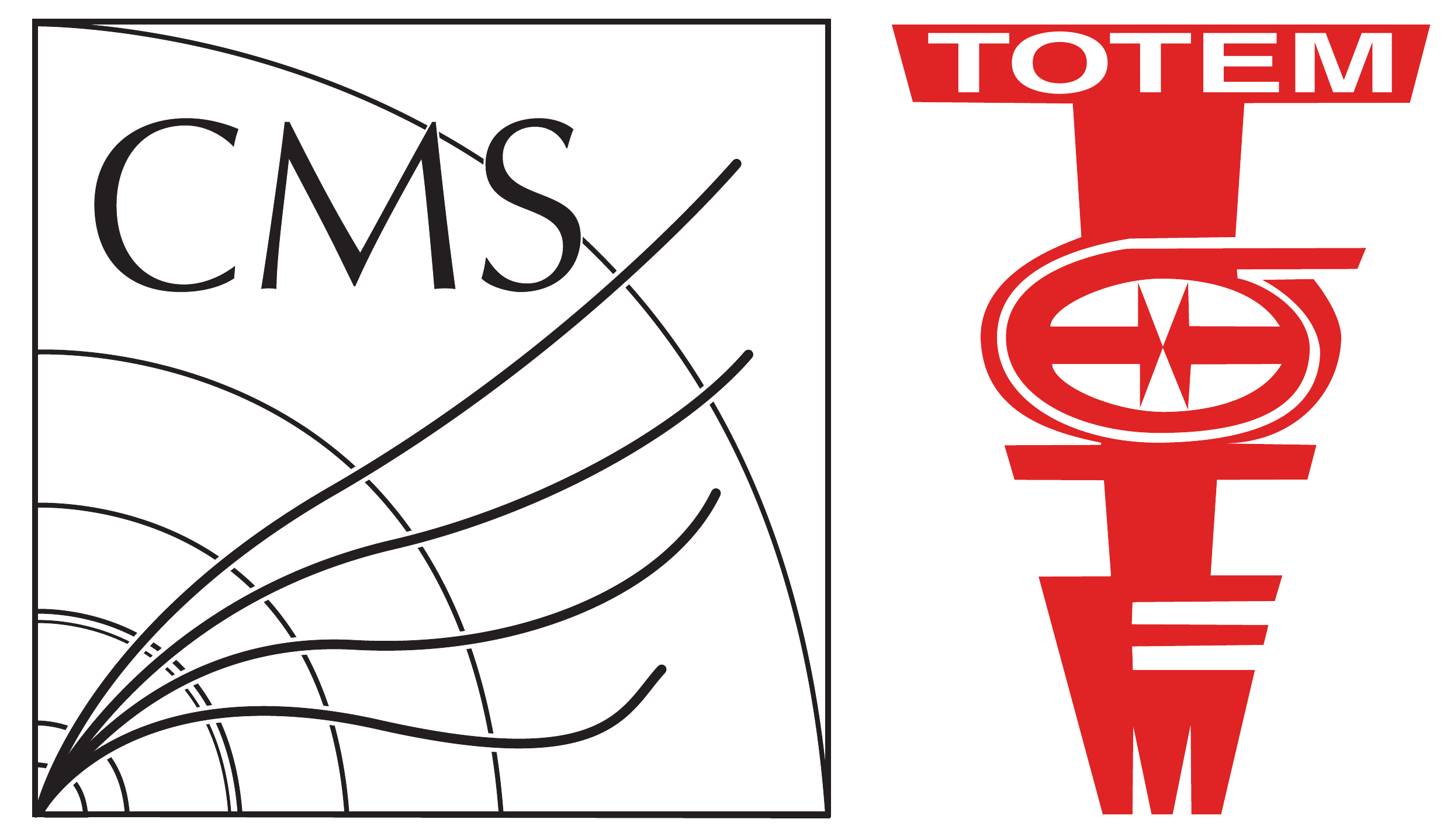}}
\renewcommand{\cmsTag}{CMS-\cmsNUMBER\\&TOTEM-2020-001\\}
\renewcommand{\cmsPubBlock}{\begin{tabular}[t]{@{}r@{}l}&CMS \cmsSTYLE\
\cmsNUMBER\\&TOTEM-2019-001\\\end{tabular}}
\renewcommand{\appMsg}{See appendices A and B for lists of collaboration members.}
\renewcommand{\cmsCopyright}{\copyright\,\the\year\ CERN for the benefit of the CMS and TOTEM Collaborations.}

\newlength\cmsTabSkip\setlength{\cmsTabSkip}{1ex}
\ifthenelse{\boolean{cms@external}}{\providecommand{\cmsTable}[1]{\resizebox{\columnwidth}{!}{#1}}}{\providecommand{\cmsTable}{}}
\providecommand{\tolerancePar}[1]{\tolerance=800 {#1}\par}
\ifthenelse{\boolean{cms@external}}{\providecommand{\clearPageSingle}{}}{\providecommand{\clearPageSingle}{\clearpage}}

\newcommand{\Ppom}{\ensuremath{\mathrm{I}\!\mathrm{P}}}
\newcommand{\Preg}{\ensuremath{\mathrm{I}\!\mathrm{R}}}
\newcommand{\cf}{\mbox{cf.}\xspace}
\newcommand{\PYTHIAvi}{{\textsc{pythia6}}\xspace}
\newcommand{\PYTHIAviii}{{\textsc{pythia8}}\xspace}
\newcommand{\HERWIGvi}{{\textsc{herwig6}}\xspace}
\newcommand{\POMWIG}{{\textsc{pomwig}}\xspace}
\newcommand{\hyph}{-\penalty0\hskip0pt\relax}

\title{Measurement of single-diffractive dijet production in proton-proton collisions at \texorpdfstring{$\sqrt{s} = 8\TeV$}{sqrt(s) = 8 TeV} with the CMS and TOTEM experiments}
\titlerunning{Single-diffractive dijet production in proton-proton collisions at \texorpdfstring{$\sqrt{s} = 8\TeV$}{sqrt(s) = 8 TeV}}
\author{The CMS and TOTEM Collaborations}

\date{\today}

\abstract{
Measurements are presented of the single-diffractive dijet cross section and the diffractive cross section as a function of the proton fractional momentum loss $\xi$ and the four-momentum transfer squared $t$.
Both processes $\Pp\Pp \to \Pp\PX$ and $\Pp\Pp \to \PX\Pp$, \ie with the proton scattering to either side of the interaction point, are measured, where $\PX$ includes at least two jets; the results of the two processes are averaged.
The analyses are based on data collected simultaneously with the CMS and TOTEM detectors at the LHC in proton-proton collisions at $\sqrt{s} = 8\TeV$ during a dedicated run with $\beta^{\ast} = 90\unit{m}$ at low instantaneous luminosity and correspond to an integrated luminosity of $37.5\nbinv$.
The single-diffractive dijet cross section $\sigma^{\Pp\PX}_{\mathrm{jj}}$, in the kinematic region $\xi < 0.1$, $0.03 < \abs{t} < 1\GeV^2$, with at least two jets with transverse momentum $\pt > 40\GeV$, and pseudorapidity $\abs{\eta} < 4.4$, is $21.7 \pm 0.9\stat \,^{+3.0}_{-3.3}\syst \pm 0.9\lum \unit{nb}$.
The ratio of the single-diffractive to inclusive dijet yields, normalised per unit of $\xi$, is presented as a function of $x$, the longitudinal momentum fraction of the proton carried by the struck parton.
The ratio in the kinematic region defined above, for $x$ values in the range $-2.9 \leq \log_{10} x \leq -1.6$, is $R = (\sigma^{\Pp\PX}_{\mathrm{jj}}/\Delta\xi)/\sigma_{\mathrm{jj}} = 0.025 \pm 0.001\stat \pm 0.003\syst$, where $\sigma^{\Pp\PX}_{\mathrm{jj}}$ and $\sigma_{\mathrm{jj}}$ are the single-diffractive and inclusive dijet cross sections, respectively.

The results are compared with predictions from models of diffractive and nondiffractive interactions.
Monte Carlo predictions based on the HERA diffractive parton distribution functions agree well with the data when corrected for the effect of soft rescattering between the spectator partons.
\vspace{12pt}
\newline
\textit{We dedicate this paper to the memory of our colleague and friend Sasha Proskuryakov, who started this analysis but passed away before it was completed. His contribution to the study of diffractive processes at CMS is invaluable.}
}

\hypersetup{%
pdfauthor={CMS Collaboration},%
pdftitle={Measurement of single-diffractive dijet production in proton-proton collisions at sqrt(s) = 8 TeV with the CMS and TOTEM experiments},%
pdfsubject={CMS-TOTEM diffractive and exclusive physics},%
pdfkeywords={CMS, TOTEM, physics, diffraction, rapidity gaps}}

\maketitle

\section{Introduction}
\label{section:introduction}

In proton-proton ($\Pp\Pp$) collisions a significant fraction of the total cross section is attributed to diffractive processes.
Diffractive events are characterised by at least one of the two incoming protons emerging from the interaction intact or excited into a low-mass state, with only a small energy loss.
These processes can be explained by the exchange of a virtual object, the so-called Pomeron,
with the vacuum quantum numbers~\cite{Collins:1977jy};
no hadrons are therefore produced in a large rapidity range adjacent to the scattered proton, yielding a so-called large rapidity gap (LRG).
A subleading exchange of Reggeons, as opposed to a Pomeron, also contributes to diffractive scattering, especially for large values of the proton fractional momentum loss $\xi$, and is required to describe diffractive data~\cite{diff_desy1,Aktas:2006hx,diff_desy2, diff_desy3}.
While Pomerons mainly consist of gluons, Reggeons are mesons composed of a quark-antiquark pair.

Hard diffraction has been studied in hadron-hadron collisions at the SPS at CERN~\cite{diff_cern}, the Tevatron at Fermilab~\cite{diff_fermi1, diff_fermi2, diff_fermi3, diff_fermi4, ratio_cdf}, the CERN LHC~\cite{Chatrchyan:2012vc, Aad:2015xis}, and in electron-proton ($\Pe\Pp$) collisions at the HERA collider at DESY~\cite{diff_desy1, Aktas:2006hx, diff_desy2, diff_desy3, diff_desy4}.
Hard diffractive processes can be described in terms of
the convolution of diffractive parton distribution functions (dPDFs) and hard scattering
cross sections, which can be calculated in perturbative quantum chromodynamics (pQCD).
The dPDFs have been determined by the HERA experiments~\cite{diff_desy1, diff_desy2, diff_desy3} by means of fits to inclusive diffractive deep inelastic scattering data.
The dPDFs have been successfully applied to describe different hard diffractive processes in $\Pe\Pp$ collisions.
This success is based on the factorisation theorem proven for $\Pe\Pp$
interactions at large $Q^2$, and on the validity of the QCD evolution equations for the dPDFs~\cite{dpdf1, dpdf2, dpdf3}.
However, in hard diffractive hadron-hadron collisions factorisation is broken because of the presence of soft rescattering between the spectator partons.
This leads to a suppression of the observed diffractive cross section in hadron-hadron collisions~\cite{Bjorken:1992er}.
The suppression factor, often called the rapidity gap survival probability ($\langle S^{2} \rangle$), is $\sim$10\% at the Tevatron energies~\cite{diff_fermi3}.

Experimentally, diffractive events can be selected either by exploiting the presence of an LRG or by measuring the scattered proton.
The latter method is superior since it gives a direct measurement of $t$, the squared four momentum transfer at the proton vertex,
and suppresses the contribution from events in which the proton dissociates into a low-mass state.
The CMS Collaboration has previously reported a measurement of diffractive dijet production at $\sqrt{s} = 7\TeV$~\cite{Chatrchyan:2012vc} that did not include information on the scattered proton.
The ATLAS Collaboration has also measured dijet production with large rapidity gaps at $\sqrt{s} = 7\TeV$~\cite{Aad:2015xis}.

This article presents a measurement of dijet production with a forward, high longitudinal momentum proton at $\sqrt{s} = 8\TeV$.
It corresponds to the processes $\Pp\Pp \to \Pp\PX$ or $\Pp\Pp \to \PX\Pp$, \ie with the proton scattering to either side of the interaction and $\PX$ including at least two jets.
The system $\PX$ is measured in CMS and the scattered proton in the TOTEM roman pots (RPs).
This process is referred to as single-diffractive dijet production.

The single-diffractive dijet production cross section is measured as a function of $\xi$ and $t$ in the kinematic region $\xi < 0.1$ and $0.03 < \abs{t} < 1\GeV^2$, in events with at least two jets, each with transverse momentum $\pt > 40\GeV$ and pseudorapidity $\abs{\eta} < 4.4$.
The ratio of the single-diffractive to inclusive dijet cross sections is measured as a function of $x$, the longitudinal momentum fraction of the proton carried by the struck parton for $x$ values in the range $-2.9 \leq \log_{10} x \leq -1.6$.
This is the first measurement of hard diffraction with a measured proton at the LHC.

\section{The CMS and TOTEM detectors}
\label{section:cms-totem-detectors}

\tolerancePar{
The central feature of the CMS apparatus is a superconducting solenoid of 6\unit{m} internal diameter,
providing a magnetic field of 3.8\unit{T}. Within the superconducting solenoid volume are a silicon pixel and strip tracker,
a lead tungstate crystal electromagnetic calorimeter (ECAL), and a brass and scintillator hadron calorimeter (HCAL),
each composed of a barrel and two endcap sections.
Forward calorimeters extend the pseudorapidity coverage provided by the barrel and endcap detectors.
The forward hadron (HF) calorimeter uses steel as an absorber and quartz fibers as the sensitive material.
The two HFs are located 11.2\unit{m} from the interaction region, one on each end, and together they provide coverage in the range $3.0 < \abs{\eta} < 5.2$.
Muons are measured in gas-ionisation detectors embedded in the steel flux-return yoke outside the solenoid.}

When combining information from the entire detector, including that from the tracker, the jet energy resolution amounts
typically to 15\% at 10\GeV, 8\% at 100\GeV, and 4\% at 1\TeV, to be compared to about 40, 12, and 5\%, respectively, obtained when
ECAL and HCAL alone are used.
In the region $\abs{\eta} < 1.74$, the HCAL cells have widths of 0.087 in pseudorapidity and 0.087 in azimuth ($\phi$). In the $\eta$-$\phi$ plane, and for $\abs{\eta} < 1.48$, the HCAL cells map on to $5{\times}5$ arrays of ECAL crystals to form calorimeter towers projecting radially outwards from close to the nominal interaction point.
For $\abs{\eta} > 1.74$, the coverage of the towers increases progressively to a maximum of 0.174 in $\Delta \eta$ and $\Delta \phi$.
Within each tower, the energy deposits in the ECAL and HCAL cells are summed to define the calorimeter tower energies, subsequently used to provide the energies and directions of hadronic jets.

\tolerancePar{
The reconstructed vertex with the largest value of summed charged-particle track $\pt^2$ is taken to be the primary interaction vertex.
Tracks are clustered based on the $z$ coordinate of the track at the point of closest approach to the beamline. In the vertex fit, each track is assigned a weight between 0 and 1, which reflects the likelihood that it genuinely belongs to the vertex. The number of degrees of freedom in the fit is strongly correlated with the number of tracks arising from the interaction region.}

The particle-flow (PF) algorithm~\cite{Sirunyan:2017ulk} aims to reconstruct and identify each individual particle in an event with an optimised combination of information from the various elements of the CMS detector.
The energy of photons is directly obtained from the ECAL measurement,
corrected for zero-suppression effects. The energy of electrons is determined from a combination of the electron momentum at
the primary interaction vertex as determined by the tracker, the energy of the corresponding ECAL cluster, and the energy sum of
all bremsstrahlung photons spatially compatible with originating from the electron track.
The energy of muons is obtained from the curvatures of the corresponding track.
The energy of charged hadrons is determined from a combination of their momentum measured
in the tracker and the matching ECAL and HCAL energy deposits, corrected for zero-suppression effects and for the response function of the
calorimeters to hadronic showers. Finally, the energy of neutral hadrons is obtained from the corresponding corrected ECAL and HCAL energies.

Hadronic jets are clustered from these reconstructed particles using the anti-\kt algorithm~\cite{Cacciari:2008gp, Cacciari:2011ma}.
The jet momentum is determined as the vectorial sum of all PF candidate momenta in the jet,
and is found from simulation to be within 5 to 10\%
of the true momentum over the whole \pt spectrum and detector acceptance.
Jet energy corrections are derived from simulation, and are confirmed with in situ measurements of the energy balance in dijet, multijet, photon + jet, and \PZ + jet events~\cite{Khachatryan:2016kdb}.
The jet \pt resolution in the simulation is scaled upwards by around 15\% in the barrel region, 40\% in the endcaps and 20\% in the forward region to match the resolution in the data.
Additional selection criteria are applied to each event to remove spurious jet-like features originating from isolated noise patterns in some HCAL regions~\cite{Chatrchyan:2009hy}.

A more detailed description of the CMS detector, together with a definition of the coordinate system used and the relevant kinematic variables, can be found in Ref.~\cite{cms}.

The TOTEM experiment~\cite{totem1,totem2} is located at the LHC interaction point (IP) 5 together with the CMS experiment.
The RP system is the subdetector relevant for measuring scattered protons.
The RPs are movable beam pipe insertions that approach the LHC beam very closely (few\mm) to detect protons scattered at very small angles or with small $\xi$. The proton remains inside the beam pipe and its trajectory is measured by tracking detectors installed inside the RPs.
They are organised in two stations placed symmetrically around the IP; one in LHC sector 45 (positive $z$), the other in sector 56 (negative $z$).
Each station is formed by two units: near (215\unit{m} from the IP) and far (220\unit{m} from the IP).
Each unit includes three RPs: one approaching the beam from the top, one from the bottom and one horizontally.
Each RP hosts a stack of 10 silicon strip sensors (pitch 66\mum) with a strongly reduced insensitive region at the edge facing the beam (few tens of \mum).
Five of these planes are oriented with the silicon strips at a $+45^{\circ}$ angle with respect to the bottom of the RP and the other five have the strips at a $-45^{\circ}$ angle.

The beam optics relates the proton kinematics at the IP and at the RP location.
A proton emerging from the interaction vertex ($x^{\ast}$, $y^{\ast}$) at horizontal and vertical angles $\theta_x^{\ast}$ and $\theta_y^{\ast}$, with a fractional momentum loss $\xi$, is transported along the outgoing beam through the LHC magnets. It arrives at the RPs at the transverse position:
\begin{linenomath}
\begin{equation}
  \begin{aligned}
  x (z_{\mathrm{RP}}) & = L_x (z_{\mathrm{RP}})\, \theta_x^{\ast} + v_x (z_{\mathrm{RP}}) \, x^{\ast} - D_x (z_{\mathrm{RP}})\, \xi , \\
  y (z_{\mathrm{RP}}) & = L_y (z_{\mathrm{RP}})\, \theta_y^{\ast} + v_y (z_{\mathrm{RP}}) \, y^{\ast} - D_y (z_{\mathrm{RP}})\, \xi ,
  \end{aligned}
\label{eq:optics}
\end{equation}
\end{linenomath}
relative to the beam centre.
This position is determined by the optical functions, characterising the transport of protons in the beamline and controlled via the LHC magnet currents.
The effective length $L_{x, y} (z)$, magnification $v_{x, y} (z)$ and horizontal dispersion $D_{x} (z)$ quantify the sensitivity of the measured proton position to the scattering angle, vertex position, and fractional momentum loss, respectively. The dispersion in the vertical plane, $D_y$, is nominally zero.

For the present measurement, a special beam optical setup with $\beta^{\ast} = 90\unit{m}$ was used, where $\beta^{\ast}$ is the value of the amplitude function of the beam at the IP.
This optical setup features parallel-to-point focussing ($v_y \sim 0$) and large $L_y$, making $y$ at RP directly proportional to $\theta_y^{\ast}$, and an almost vanishing $L_x$ and $v_x$, implying that any horizontal displacement at the RP is approximately proportional to $\xi$.
Protons can hence be measured with large detector acceptance in the vertical RPs that approach the beam from the top and bottom.

To reduce the impact of imperfect knowledge of the optical setup, a calibration procedure~\cite{lhc_optics} has been applied.
This method uses elastic scattering events and various proton observables to determine fine corrections to the optical functions presented in Eq.~(\ref{eq:optics}).
For the RP alignment, a three-step procedure~\cite{totem2} has been applied: beam-based alignment prior to the run (as for the LHC collimators) followed by two offline steps.
First, track-based alignment for the relative positions among RPs, and second, alignment with elastic events for the absolute position with respect to the beam.
The final uncertainties per unit (common for top and bottom RPs) are: 2\mum (horizontal shift), 100\mum (vertical shift), and 0.2\unit{mrad} (rotation about the beam axis).

The kinematic variables ($\xi$, $\theta_x^{\ast}$, $\theta_y^{\ast}$ as well as $t$) are reconstructed with the use of parametrised proton transport functions~\cite{totem2}.
The values of the optical functions vary with $\xi$, an effect that is taken into account by the optics parametrisation.
The details of the reconstruction algorithms and optics parametrisation are discussed in Refs.~\cite{totem2,Niewiadomski:2008zz}.
The momentum loss reconstruction depends mostly on the horizontal dispersion, which is determined with a precision better than 10\%.
The scattering angle resolution depends mainly on the angular beam divergence and in the horizontal plane also on the detector resolution, whereas the momentum loss resolution depends mainly on the optics~\cite{totem3}.
The $\xi$ resolution is about $\sigma$($\xi$) = 0.7\% and the $\theta_y^{\ast}$ and the $\theta_x^{\ast}$ resolutions 2.4\unit{$\mu$rad} and 25\unit{$\mu$rad}, respectively.

\section{Event kinematics}
\label{section:kinematics}

\begin{figure}[hbtp]
  \centering
    \includegraphics[width=\cmsFigWidthSmall]{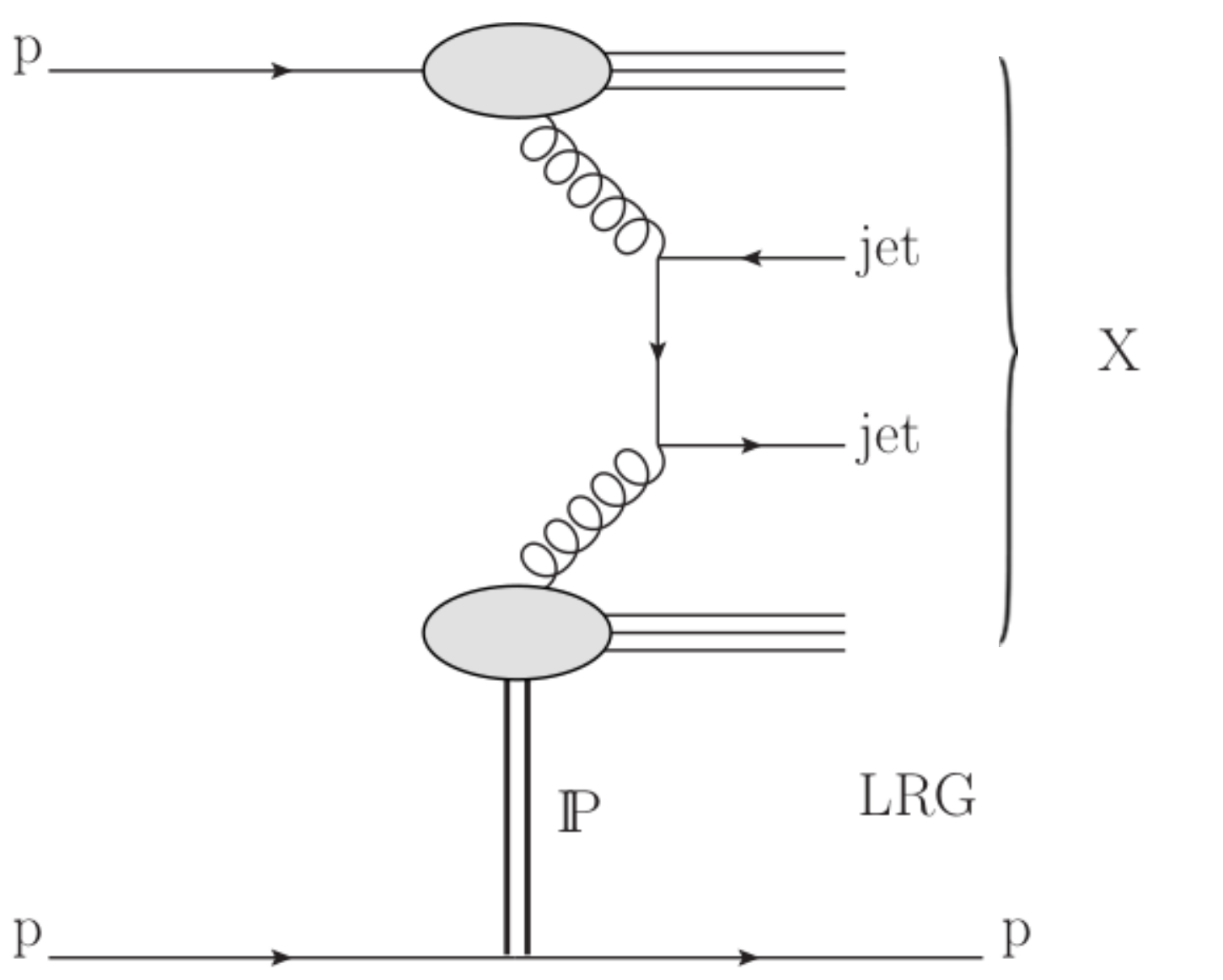}
     \caption{
      Schematic diagram of single-diffractive dijet production.
      The exchange of a virtual object with the vacuum quantum numbers (\ie a Pomeron) is indicated by the symbol $\textrm{I}\!\textrm{P}$.
      The diagram shows an example of the $\Pg\Pg \to \text{dijet}$ hard scattering process;
      the $\Pq\Pq$ and $\Pg\Pq$ initial states also contribute.
     }
    \label{fig:dijet_graph}
\end{figure}

\tolerancePar{
Figure \ref{fig:dijet_graph} shows a schematic diagram of the single-diffractive
reaction $\Pp\Pp \to \PX\Pp$ with $\PX$ including two high-$\pt$ jets.
Single-diffractive dijet production is characterised by the presence of a high-energy proton, which escapes undetected by the CMS detector, and the system $\PX$, which contains high-$\pt$ jets, separated from the proton by an LRG.}

The proton is scattered at small angles, has small fractional
momentum loss $\xi = 1 - {\abs{\mathbf{p}_f}} / {\abs{\mathbf{p}_i}}$,
and small absolute value of the 4-momentum transfer squared $t = (p_f - p_i)^2$, where $p_i$ and $p_f$ are the four-momenta of the incoming and outgoing protons, respectively.
The scattered proton does not leave the beam pipe and can only be detected by using the TOTEM RP detectors, which make a direct measurement of $t$ and $\xi$ (hereafter referred to as $\xi_{\text{\tiny TOTEM}}$).

If only CMS information is used, as in Ref.~\cite{Chatrchyan:2012vc}, $\xi$ can be estimated only from the energies and longitudinal momenta of the particles measured in CMS:
\begin{equation}
  {\xi}_{\text{\tiny CMS}}^{\pm}= \frac{ \sum_i \left( E^i\pm p_z^i \right) }{\sqrt{s}},
  \label{eq:xi_cms}
\end{equation}
where the sum is carried out with PF objects.
The positive (negative) sign corresponds to the scattered proton moving towards the positive (negative) $z$ direction. In this case, $t$ cannot be measured.

The combination of the limited CMS pseudorapidity coverage ($\abs{\eta} < 5$) and the detector inefficiency causes $\xi_{\text{\tiny CMS}}$ to be smaller than $\xi_{\text{\tiny TOTEM}}$ in general, \ie $\xi_{\text{\tiny CMS}} - \xi_{\text{\tiny TOTEM}} \leq 0$.
However, the limited detector resolution may cause $\xi_{\text{\tiny CMS}}$ to be larger than $\xi_{\text{\tiny TOTEM}}$.

The momentum fraction of the partons initiating the hard scattering, $x^{+}$ and $x^{-}$, can be estimated from the energies and longitudinal momenta of the measured jets as:
\begin{equation}
  x^{\pm}= \frac{ \sum_{\text{\tiny jets}}\left( E^{\text{\tiny jet}} \pm p_z^\text{\tiny jet} \right) }{\sqrt{s}},
  \label{eq:x_jets_cms}
\end{equation}
where the sum is carried out over the two highest transverse momentum jets in the event, and an additional third jet, if present.
The positive (negative) sign corresponds to the incoming proton moving towards the positive (negative) $z$ direction.

Finally, the fraction $\beta$ of the Pomeron momentum carried by the interacting parton is measured from the values of $x^{\pm}$ and $\xi_{\text{\tiny TOTEM}}$ as $\beta = x^{\pm}/\xi_{\text{\tiny TOTEM}}$.

\section{Data samples}
\label{section:data-mc-samples}

The data were collected in July 2012 during a dedicated run with low probability ($\sim$6--10\%) of overlapping $\Pp\Pp$ interactions in the same bunch crossing (pileup) and a nonstandard $\beta^*=90\unit{m}$ beam optics configuration.
These data correspond to an integrated luminosity of $\lumi = 37.5\nbinv$.
Events are selected by trigger signals that are delivered simultaneously to the CMS and TOTEM detectors.
The first level of the CMS trigger system (L1) is used. The L1 signal is propagated to the TOTEM electronics to enable the simultaneous readout of the CMS and TOTEM subdetectors.
The CMS orbit-counter reset signal, delivered to the TOTEM electronics at the start of the run, assures the time synchronisation of the two experiments.
The CMS and the TOTEM events are combined offline based on the LHC orbit and bunch numbers.

\section{Monte Carlo simulation}
\label{section:mc-simulation-acceptance}

The simulation of nondiffractive dijet events is performed with the \PYTHIAvi (version 6.422) \cite{pythia6}, \PYTHIAviii~(version 8.153)~\cite{pythia8}, and \HERWIGvi~\cite{herwig6} Monte Carlo (MC) event generators.
The underlying event is simulated in \PYTHIAvi with tune Z2*~\cite{Field:2010bc} and in \PYTHIAviii with tunes 4C~\cite{Corke:2010yf}, CUETP8M1, and CUETP8S1~\cite{Khachatryan:2015pea}.

Single-diffractive dijet events are simulated with the \PYTHIAviii and \POMWIG~(version 2.0)~\cite{pomwig} generators.
Hard diffraction is simulated in \PYTHIAviii using an inclusive diffraction model, where both low- and high-mass systems are generated~\cite{Navin:2010kk}.
High-mass diffraction is simulated using a perturbative description. Pomeron parton densities are introduced and the diffractive process is modelled as a proton-Pomeron scattering at a reduced centre-of-mass energy.
The default generator settings are used, including that for the proton-Pomeron total cross section.
Multiparton interactions (MPI) are included within the proton-Pomeron system to provide cross sections for parton-parton interactions.
In this model, the presence of secondary interactions does not lead to a suppression of the visible diffractive cross section.

Additionally, \PYTHIAviii implements a model to simulate hard-diffractive events based on a direct application of dPDFs, and a dynamical description of the rapidity gap survival probability in diffractive hadron-hadron interactions~\cite{Rasmussen:2015qgr}.
In this model an event is classified as diffractive only when no MPI are generated.
We refer to this implementation as the dynamic gap (DG) model.
Single-diffractive dijet events using the inclusive diffraction model are simulated with \PYTHIAviii, tunes 4C and CUETP8M1.
The simulation of diffractive dijet events using the DG model is performed with \PYTHIAviii~version 8.223~\cite{Rasmussen:2015qgr} with the underlying event tune CUETP8M1.
These \PYTHIAviii tunes give a fair description of the charged-particle pseudorapidity and \pt distributions in a sample with a large fraction of single-diffractive inelastic events~\cite{Chatrchyan:2014qka,Khachatryan:2015pea,Sirunyan:2018zdc}.

The \POMWIG generator is based on \HERWIGvi and implements dPDFs to simulate hard-diffractive processes.
The simulation uses dPDFs from a fit to deep inelastic scattering data~(H1 fit B~\cite{diff_desy1}).
The \POMWIG generator uses a next-to-leading order dPDF fit, whereas \PYTHIAviii uses a leading order dPDF fit.
When using \POMWIG, a constant factor $\langle S^{2} \rangle = 7.4\%$ is applied to account for the rapidity gap survival probability leading to the suppression of the diffractive cross section.
This value is calculated from the ratio of the measured diffractive cross section and the prediction from \POMWIG, as described in Section~\ref{section:results-dsigmadtdxi}.
Both Pomeron and Reggeon exchange contributions are generated. Reggeon exchange is not simulated in \PYTHIAviii.

To improve the description of the data by the MC samples, correction factors are applied event-by-event as a function of $\beta$, by a reweighting procedure.
The correction modifies the event distribution as a function of $\beta$ by up to 40\%, and the $\log_{10}x$ and $\xi$ distributions by as much as 30\% and 8\%, respectively. The correction has a negligible effect on the $t$ distribution.  

The generated events are processed through the simulation of the CMS detector, based on \GEANTfour~\cite{Agostinelli:2002hh},
and reconstructed in the same manner as the data.
The acceptance and resolution of the TOTEM RP detectors are parametrised as a function of the proton kinematics, as discussed below.
All samples are simulated without pileup.

\subsection{Roman pot detectors acceptance and resolution}

The proton path from the IP to the TOTEM RPs is calculated using a parametrisation of the LHC optics~\cite{lhc_optics}.
To obtain a realistic simulation of the scattered proton, the following procedure is used:

\begin{itemize}

    \item \textit{Proton transport:}
    The simulation of the RP detectors acceptance is parametrised in terms of the vertex position, the proton scattering angles at the vertex $\theta_x^{\ast}$ and $\theta_y^{\ast}$, and $\xi$.
    The incident beam energy spread and beam divergence are also simulated~\cite{totem3}.

    \item \textit{Reconstruction of $t$ and $\xi$:}
    The detector-level distributions of $t$ and $\xi$ are obtained from the scattering angles $\theta_x^{\ast}$ and $\theta_y^{\ast}$,
    where the correlation between the $\xi$ and $\theta_x^{\ast}$ uncertainties is taken into account~\cite{totem2}.
    The generated values of $\theta_x^{\ast}$ and $\theta_y^{\ast}$ are spread by $25\unit{$\mu$rad}$ and $2.4\unit{$\mu$rad}$, respectively.
    These values include the effects of detector resolution, as well as those of the beam optics and the beam divergence.

    \item \textit{Proton reconstruction inefficiency:}
    The track reconstruction in the RPs may fail for several reasons: inefficiency of the silicon sensors,
    interaction of the proton with the RP mechanics, or the simultaneous presence of a beam halo particle or a proton from a pileup interaction.
    The silicon strips of the detectors in an RP are oriented in two orthogonal directions;
    this allows for good rejection of inclined background tracks, but makes it very difficult to reconstruct more than one track almost parallel to the beam direction~\cite{totem2}.
    These uncorrelated inefficiencies are evaluated from elastic scattering data~\cite{totem3}, and amount to $\sim$6\%.
    To correct for this, an extra normalisation factor is applied, obtained separately for protons traversing the RPs on either side of the IP.

\end{itemize}

\section{Event selection}
\label{section:event-selection}

Dijet events are selected online by requiring at least two jets with $\pt > 20\GeV$~\cite{Khachatryan:2016bia}.
The efficiency of this trigger selection is estimated with a sample of minimum bias events, \ie events collected with a loose trigger intended to select inelastic collisions with as little bias as possible, and containing a leading jet with $\pt$, as reconstructed offline, of at least 40\GeV.
The fraction of dijet events accepted by the trigger is calculated as a function of the subleading jet $\pt$. The efficiency is above $94\%$ for $\pt > 40\GeV$.

The offline selection requires at least two jets with $\pt > 40\GeV$ and $\abs{\eta} < 4.4$.
Jets are reconstructed from PF objects with the anti-\kt algorithm with a distance parameter $R=0.5$.
The reconstructed jet energy is corrected with the procedure described in Ref.~\cite{Khachatryan:2016kdb}.
The parton momentum fractions $x^{+}$ and $x^{-}$ are reconstructed using Eq.~(\ref{eq:x_jets_cms}) from the two highest transverse momentum jets and an additional third jet, if present.
The latter is selected with $\pt > 20\GeV$.
In addition, the selection requires at least one reconstructed primary interaction vertex and at least one reconstructed proton track in the RP stations.
The fit of the reconstructed vertex is required to have more than four degrees of freedom.

Events with protons in the RP stations on both sides are rejected if their kinematics are consistent with those of elastic scattering.
Elastic scattering events, which are present in the data sample because of pileup, are identified by the presence of two proton tracks in opposite directions, in a diagonal configuration: the protons traverse the two top RPs in sector 45 and the two bottom RPs in sector 56, or vice versa. The horizontal and vertical scattering angles are required to match within the measured resolutions.
These requirements are similar to those described in Ref.~\cite{totem3}.

To avoid detector edges with rapidly varying efficiency or acceptance, as well as regions dominated by secondary particles produced by aperture limitations in the beamline upstream of the RPs,
proton track candidates are selected if the corresponding hit coordinates on the RP stations satisfy the following
fiducial requirements: $0 < x < 7\mm$ and $8.4 < \abs{y} < 27\mm$,
where $x$ and $y$ indicate the horizontal and vertical coordinates of the hit with respect to the beam.

To suppress background from secondary particles and pileup in the RPs, the reconstructed proton track is selected if it is associated to one track element in both top or both bottom RPs on a given side.
The kinematic requirements $0.03 < \abs{t} < 1.0\GeV^2$ and $0 < \xi_{\text{\tiny TOTEM}} < 0.1$ are then applied.

For signal events, one expects $\xi_{\text{\tiny CMS}}$ to be smaller than $\xi_{\text{\tiny TOTEM}}$, \ie $\xi_{\text{\tiny CMS}} - \xi_{\text{\tiny TOTEM}} \leq 0$ (as discussed in Section~\ref{section:kinematics}).
This selection is imposed to suppress the contribution of pileup and beam halo events, in which the proton is uncorrelated with the hadronic final state $\PX$ measured in the CMS detector.
Roughly $6\%$ of signal events are rejected by this requirement, as estimated from a simulation of single-diffractive dijet production.

Table~\ref{table:selection} shows the number of events passing each selection.
The number of events with the proton detected in the RPs in sector 45~(56) after all the selections is 368~(420).

A difference in the yields for events with a proton in sector 45 and 56 could notably arise from different background contributions, which is discussed in Section~\ref{section:background}. Both an imperfect knowledge of the optical functions, especially the horizontal dispersion, discussed in Section~\ref{section:results-systematics}, and statistical fluctuations of the two mostly independent event samples contribute to the difference. 

\begin{table}[hbtp]
 \centering
  \topcaption{Number of events after each selection.}
  \label{table:selection}
  \cmsTable{
  \begin{tabular}{ccc}
   \hline
   Selection & Sector 45 & Sector 56 \\
   \hline
   At least 2 jets ($\pt > 40\GeV$, $\abs{\eta} < 4.4$) & \multicolumn{2}{ c }{427689}\\
   Elastic scattering veto & \multicolumn{2}{ c }{405112}\\
   Reconstructed proton & \multicolumn{2}{ c }{9530}\\
   RP and fiducial region & 2137  & 3033\\
   $0.03 < \abs{t} < 1.0\GeV^2$, $0 < \xi_{\text{\tiny TOTEM}} < 0.1$ & 1393 & 1806 \\
   $\xi_{\text{\tiny CMS}} - \xi_{\text{\tiny TOTEM}}\leq0$ & 368 & 420 \\
   \hline
  \end{tabular}
  }
\end{table}

\section{Background}
\label{section:background}

The main background is due to the overlap of a $\Pp\Pp$ collision in the CMS detector and an additional track in the RP stations, originating from either a beam halo particle or an outgoing proton from a pileup interaction.

Pileup and beam halo events are not simulated, but they are present in the data.
To estimate the pileup and beam halo contribution in the data, a zero bias sample consisting of events from randomly selected, nonempty LHC bunch crossings is used.
Events with a proton measured in the RP stations and with any number of reconstructed vertices are selected from the zero bias data set.
Such events are denoted by ZB in the following.

The RP information from events in the zero bias sample is added to diffractive and nondiffractive events generated with \POMWIG and \PYTHIAvi, respectively.
The mixture of MC and ZB events simulates data events in the presence of pileup and beam halo.

The \POMWIG sample is normalised assuming a rapidity gap survival probability factor of 7.4\%, as discussed in Section~\ref{section:mc-simulation-acceptance}.
The MC and ZB event mixture is then passed through the selection procedure illustrated in Section~\ref{section:event-selection},
except for the requirement ${\xi_{\text{\tiny CMS}} - \xi_{\text{\tiny TOTEM}} \leq 0}$, which is not applied.

\begin{figure*}[hbtp]
  \centering
    \includegraphics[width=\cmsFigWidth]{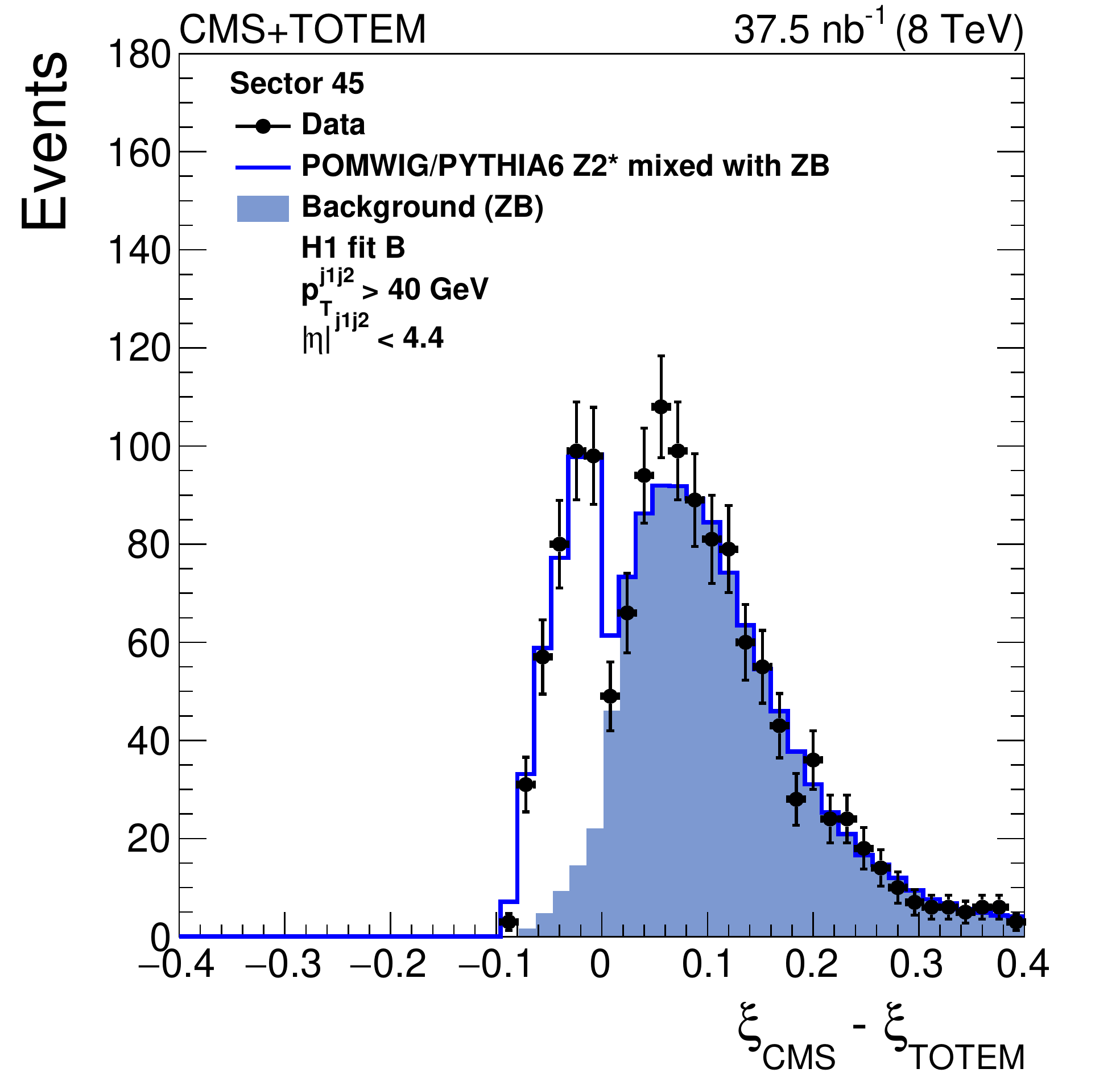}
    \includegraphics[width=\cmsFigWidth]{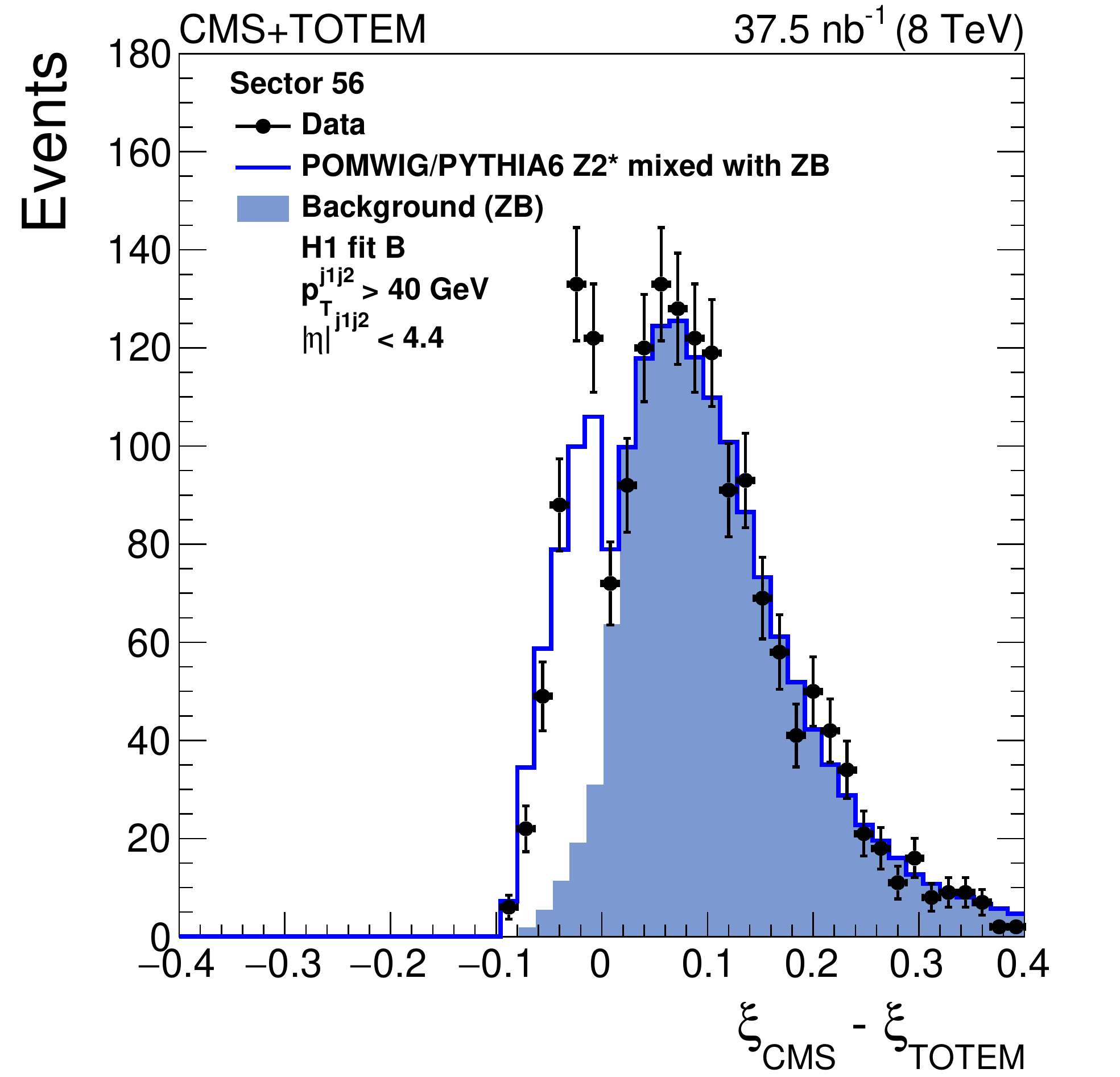}
     \caption{
      Distribution of $\xi_{\text{\tiny CMS}} - \xi_{\text{\tiny TOTEM}}$ for events with a reconstructed proton in sector 45 (left) and sector 56 (right).
      The data are indicated by solid circles.
      The blue histogram is the mixture of \POMWIG or \PYTHIAvi and zero bias (ZB) data events described in the text.
      An event with a proton measured in the RPs contributes to the open histogram (signal) if the proton originates from the MC sample, or to the filled histogram (background) if it originates from the ZB sample.
      }
    \label{fig:background:xi_cms_totem}
\end{figure*}

Such mixed events with a proton in the RPs are considered as signal if the proton originates from the MC simulated sample, or as background if it originates from the ZB sample.
If an event has a proton from both the MC sample and the ZB sample, the proton with smaller $\xi$ is chosen.
However, the probability of such a combination is small and none of these events pass all the selections.
Figure~\ref{fig:background:xi_cms_totem} shows the distribution of $\xi_{\text{\tiny CMS}} - \xi_{\text{\tiny TOTEM}}$ for the data compared to the MC+ZB event mixture.
The requirement $\xi_{\text{\tiny CMS}} - \xi_{\text{\tiny TOTEM}} \leq 0$ selects signal events and rejects the kinematically forbidden region populated by the MC+ZB background events (filled histogram).
The background distribution is normalised to the data in the $\xi_{\text{\tiny CMS}} - \xi_{\text{\tiny TOTEM}}$ region from 0.048 to 0.4, which is dominated by background events.

The background is estimated separately for events with a proton traversing the two top (top-top) or the two bottom (bottom-bottom) RPs on each side. The top-top and bottom-bottom distributions are similar.
Figure~\ref{fig:background:xi_cms_totem} shows the sum of the two contributions.

The background contribution for events with a proton detected in sector 56 (right panel of Fig.~\ref{fig:background:xi_cms_totem}) is larger than that for events with a proton detected in sector 45 (left panel of Fig.~\ref{fig:background:xi_cms_totem}).
The remaining contamination of background in the signal region is estimated to be $15.7\%$ for events in which the proton is detected in sector 45 and $16.8\%$ for those in which the proton is detected in sector 56.

\begin{figure*}[hbtp]
  \centering
    \includegraphics[width=\cmsFigWidthSingleSmall]{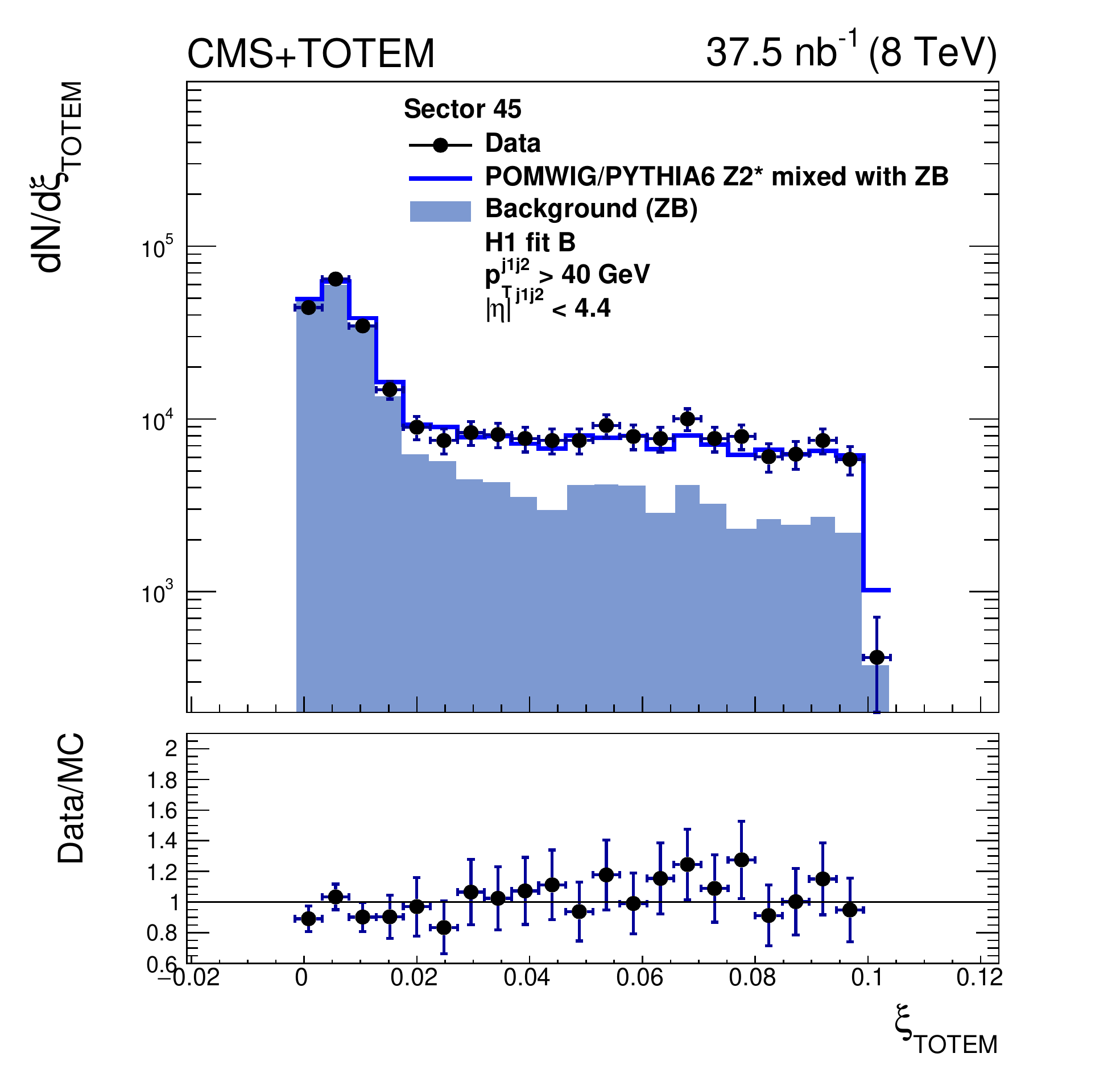}
    \includegraphics[width=\cmsFigWidthSingleSmall]{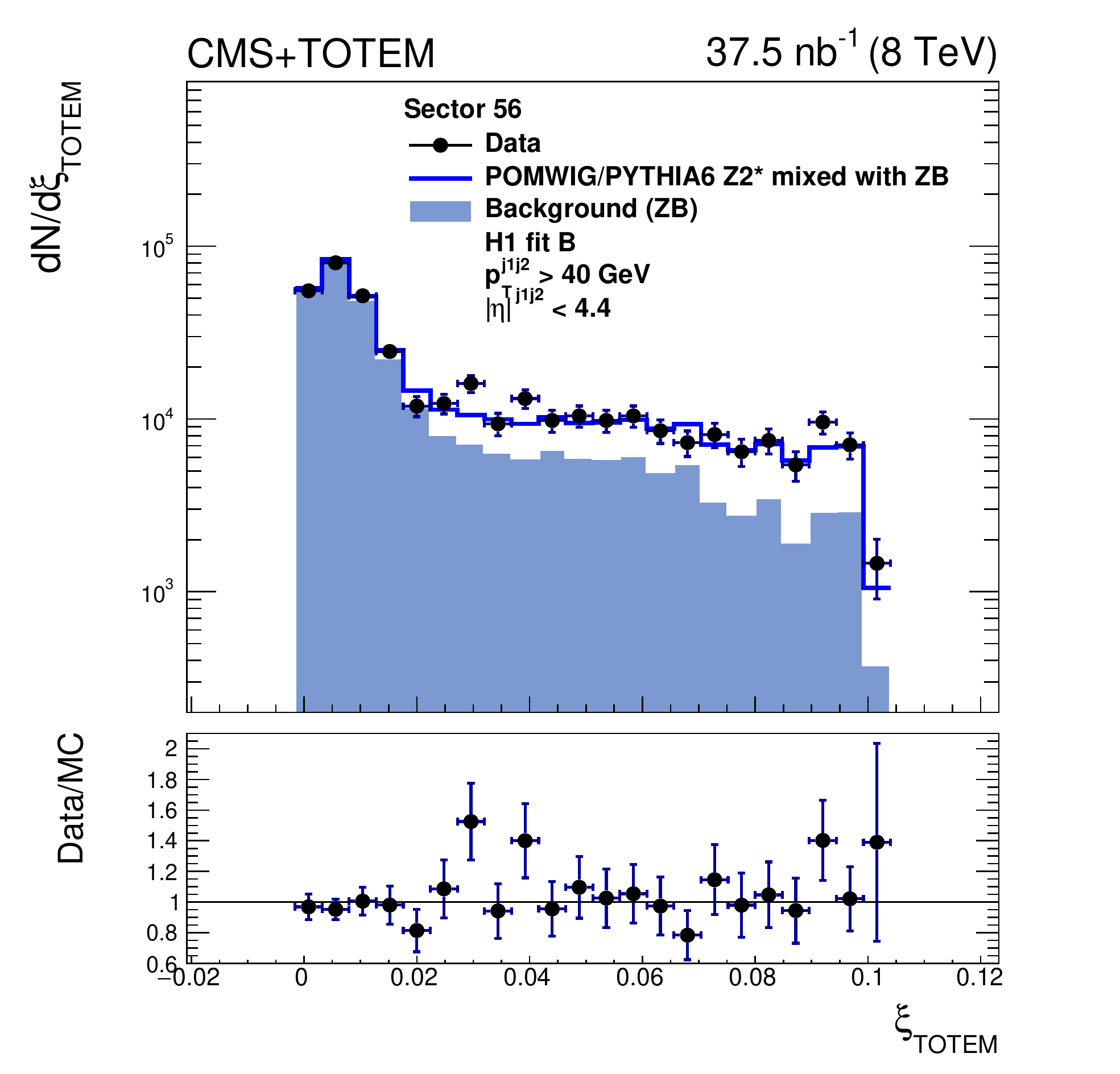}
    \includegraphics[width=\cmsFigWidthSingleSmall]{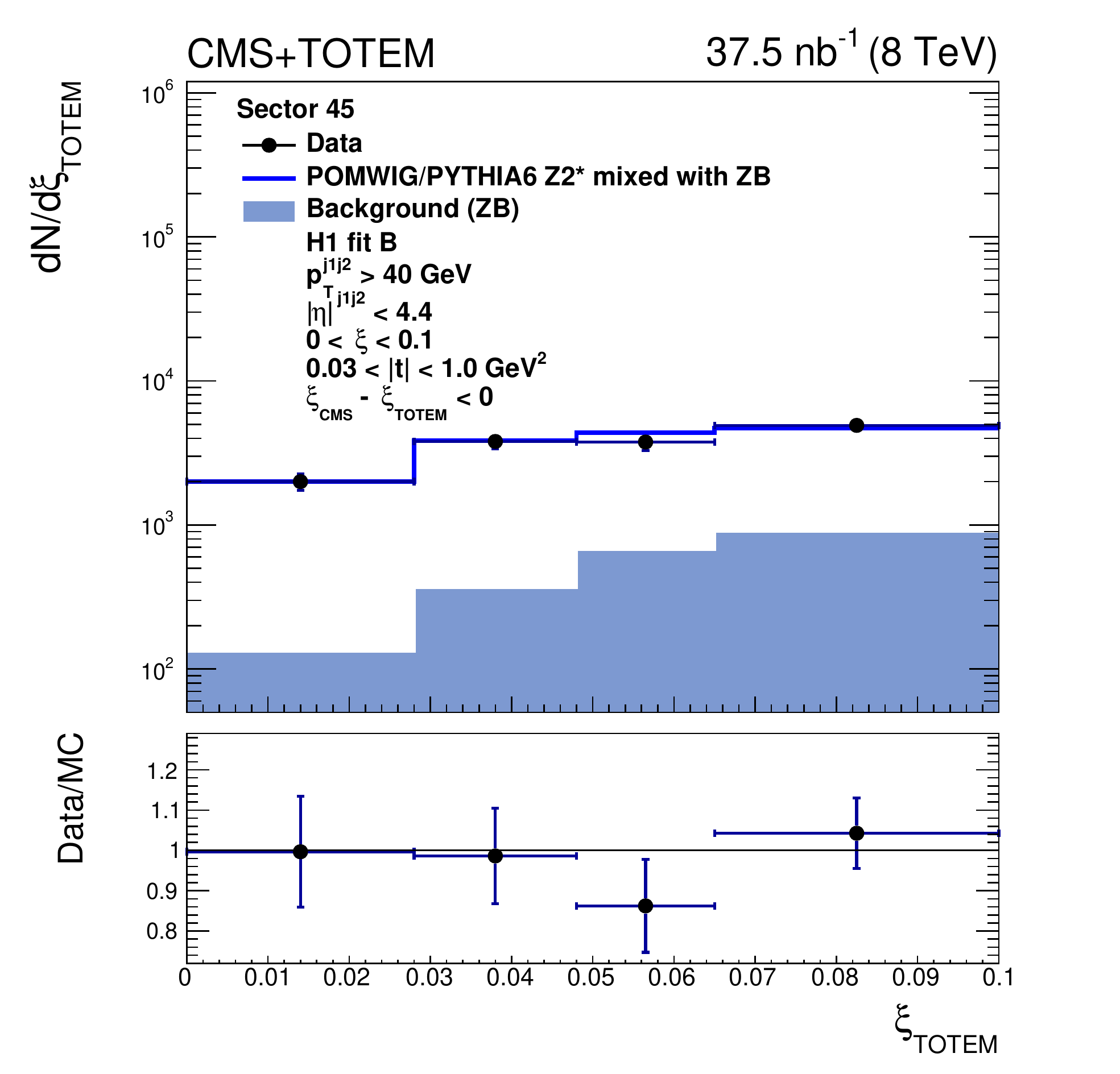}
    \includegraphics[width=\cmsFigWidthSingleSmall]{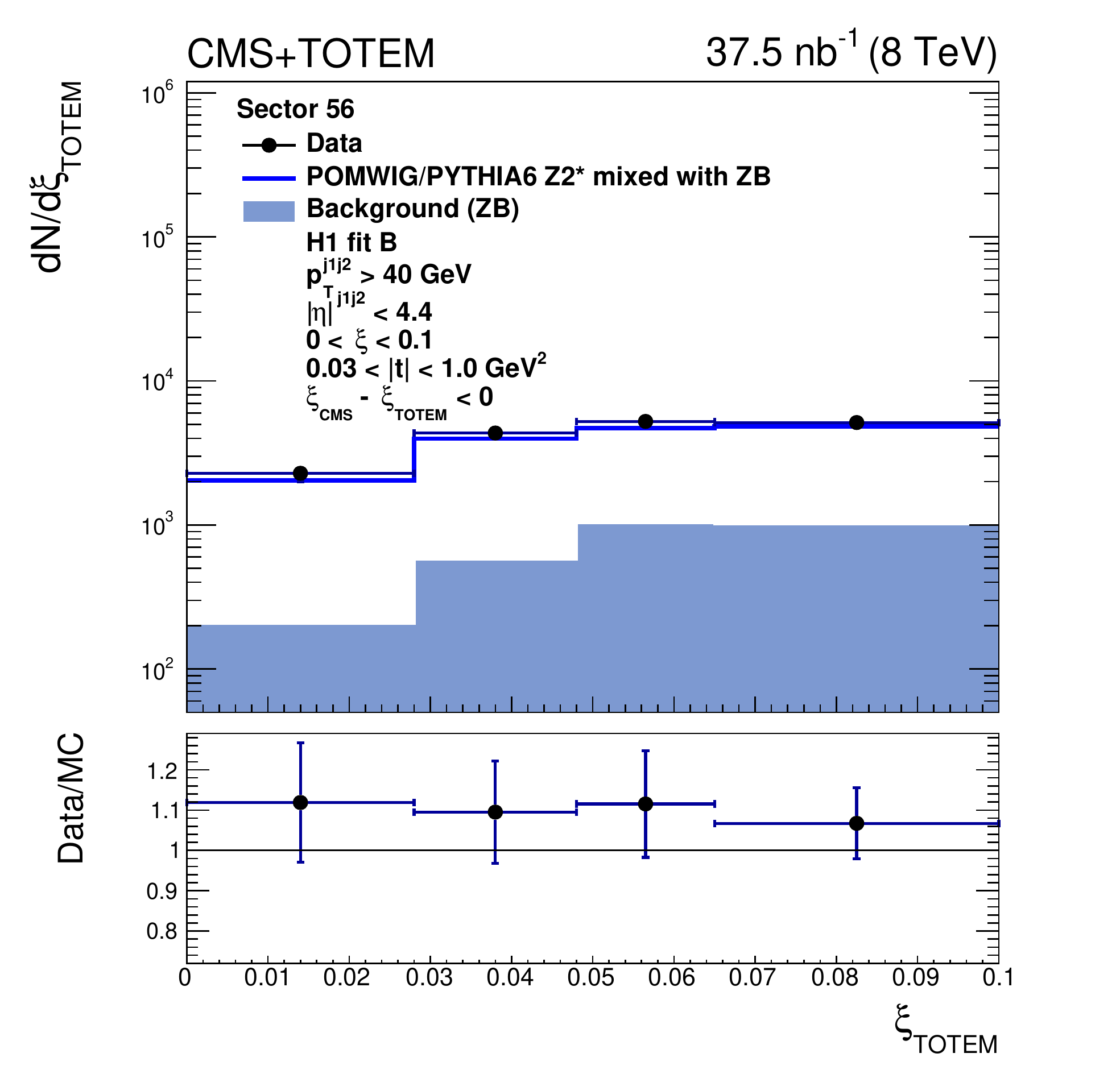}
    \includegraphics[width=\cmsFigWidthSingleSmall]{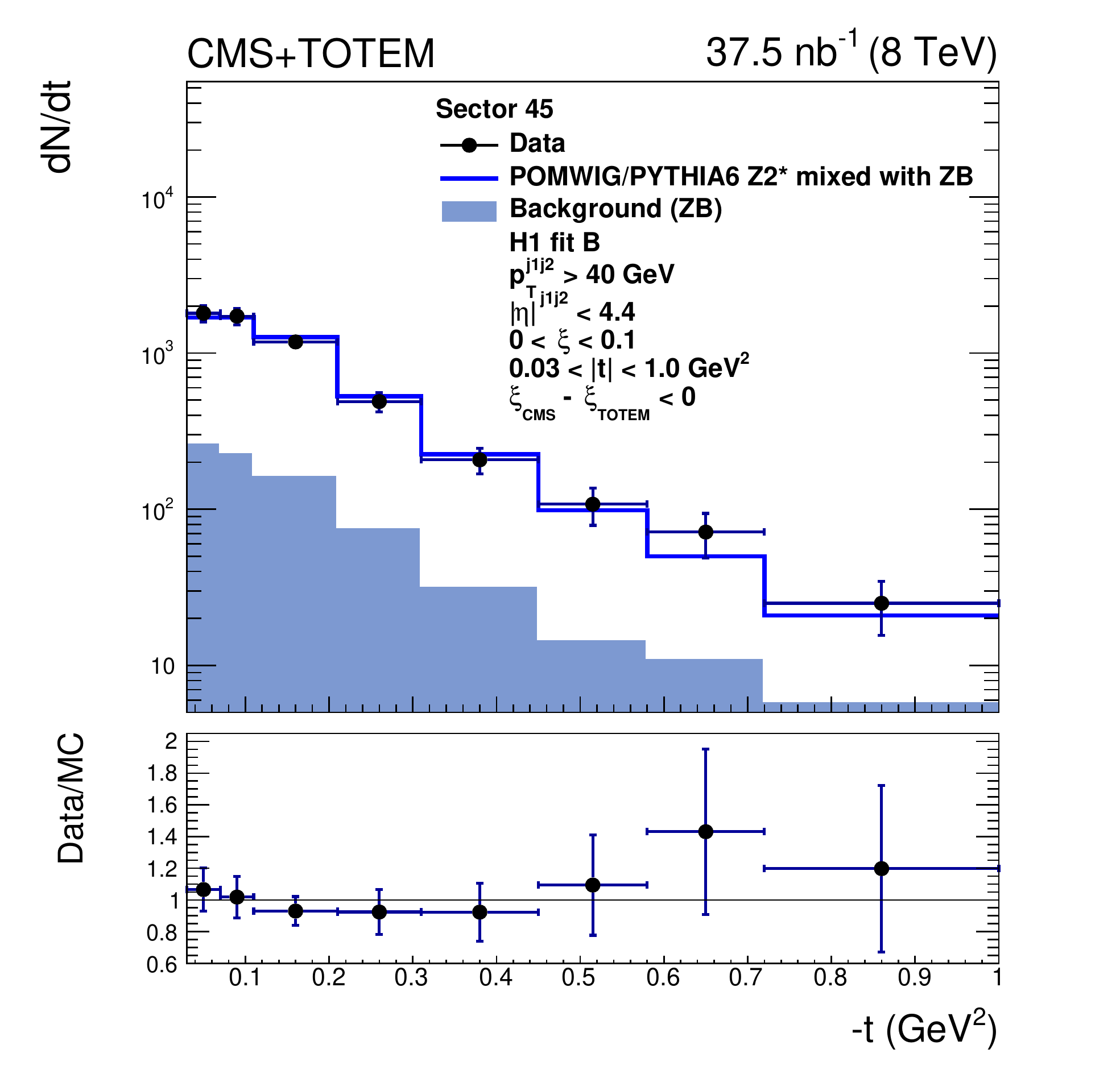}
    \includegraphics[width=\cmsFigWidthSingleSmall]{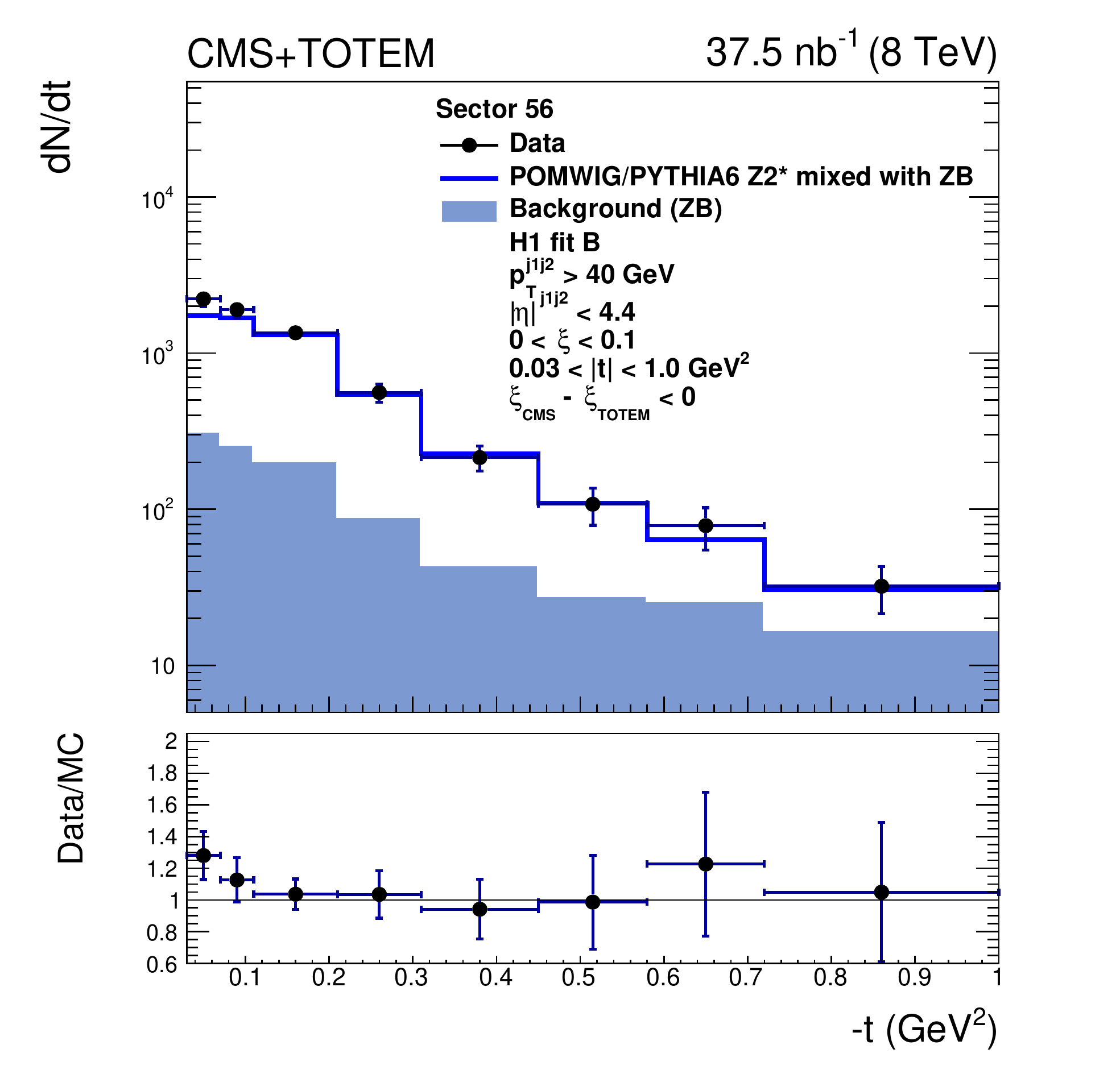}
     \caption{
      Distribution of $\xi_{\text{\tiny TOTEM}}$ before (upper) and after (middle) the $\xi_{\text{\tiny CMS}} - \xi_{\text{\tiny TOTEM}}$
      requirement and distribution of $t$ after the $\xi_{\text{\tiny CMS}} - \xi_{\text{\tiny TOTEM}}$ requirement (lower) for events in which the proton is detected in sector 45 (left) and sector 56 (right).
      The data are indicated by solid circles.
      The blue histogram is the mixture of \POMWIG or \PYTHIAvi and zero bias (ZB) data events described in the text.
      An event with the proton measured in the RPs contributes to the open histogram (signal) if the proton originates from the MC sample, or to the filled histogram (background) if it originates from the ZB sample.
      }
    \label{fig:background:xi_t_totem}
\end{figure*}

Figure~\ref{fig:background:xi_t_totem} shows the distribution of $\xi_{\text{\tiny TOTEM}}$ for the data and the MC+ZB sample, before and after the $\xi_{\text{\tiny CMS}} - \xi_{\text{\tiny TOTEM}}\leq0$ requirement, as well as the distribution of $t$, after the $\xi_{\text{\tiny CMS}} - \xi_{\text{\tiny TOTEM}}\leq0$ selection. The sum of the top-top and bottom-bottom combinations is used. The data and the MC+ZB sample are in good agreement.

An alternative method, used at HERA~\cite{diff_desy2}, takes two events randomly chosen from the data sample.
First, $\xi_{\text{\tiny CMS}}$ is sampled from events that have passed the dijet selection;
$\xi_{\text{\tiny TOTEM}}$ is then taken from events with $\xi_{\text{\tiny CMS}} > 0.12$ that have passed the event selection described in Section~\ref{section:event-selection}, except for the ${\xi_{\text{\tiny CMS}} - \xi_{\text{\tiny TOTEM}}}$ requirement, to select proton tracks considered to be mostly from background.
These two values are used to plot the $\xi_{\text{\tiny CMS}} - \xi_{\text{\tiny TOTEM}}$ distribution, which is normalised to the data in a region dominated by background.
The remaining contamination in the signal region is $\sim$19\% both for events with a proton detected in sector 45 and for those with a proton in sector 56.
The ZB method is used in this analysis.
Half the difference between the results of the two methods is taken as an estimate of the systematic uncertainty of the background subtraction procedure.

\section{Results}
\label{section:results}

In this section the measurements of the differential cross sections $\rd\sigma/{\rd}t$, $\rd\sigma/\rd\xi$, and the ratio $R(x)$ of the single-diffractive ($\sigma^{\Pp\PX}_{\mathrm{jj}}(x)$) to inclusive dijet cross sections ($\sigma_{\mathrm{jj}}(x)$) are presented.
The ratio $R(x)$, normalised per unit of $\xi$, is defined by:
\begin{equation}
 R(x)=\frac{\sigma^{\Pp\PX}_{\mathrm{jj}}(x)/\Delta\xi}{\sigma_{\mathrm{jj}}(x)} ,
 \label{eq:results:ratio}
\end{equation}
where $\Delta\xi = 0.1$.

The cross sections are calculated in the kinematic region $\xi < 0.1$, $0.03 < \abs{t} < 1\GeV^2$,
with at least two jets at a stable-particle level with $\pt > 40\GeV$ and $\abs{\eta} < 4.4$.
The ratio $R(x)$ is calculated for $x$ values in the region $-2.9\leq\log_{10} x\leq-1.6$.
In the following, the estimated background is subtracted from the number of single-diffractive dijet candidates following the procedure described in the previous section.

The differential cross sections for dijet production in bins of $t$ and $\xi$ are evaluated as:
\begin{equation}
 \begin{aligned}
   \frac{\rd\sigma^{\Pp\PX}_{\mathrm{jj}}}{{\rd}t} &= \mathcal{U}\left\{ \frac{N^{\Pp\PX}_{\mathrm{jj}}}{\mathcal{L} A_{\text{\tiny CMS-TOTEM}} {\Delta t}} \right\}, \\
   \frac{\rd\sigma^{\Pp\PX}_{\mathrm{jj}}}{\rd\xi} &= \mathcal{U}\left\{ \frac{N^{\Pp\PX}_{\mathrm{jj}}}{\mathcal{L} A_{\text{\tiny CMS-TOTEM}} {\Delta \xi}} \right\},
 \end{aligned}
 \label{eq:results:dsigmadtdxi}
\end{equation}
where $N^{\Pp\PX}_{\mathrm{jj}}$ is the measured number of single-diffractive dijet candidates per bin of the distribution after subtracting the estimated background;
${\Delta t}$ and ${\Delta \xi}$ are the bin widths, and $\mathcal{L}$ is the integrated luminosity. 
The factors $A_{\text{\tiny CMS-TOTEM}}$ indicate the acceptance of CMS and TOTEM for single-diffractive dijet events.
Unfolding corrections, represented by the symbol~$\mathcal{U}$ in Eq.~(\ref{eq:results:dsigmadtdxi}), are applied to account for the finite resolution of the reconstructed variables used in the analysis. They are evaluated with \POMWIG, \PYTHIAviii 4C and \PYTHIAviii CUETP8M1.
The results presented are the average of those obtained with the different unfolding corrections.
The measured cross sections are obtained by unfolding the data using the D'Agostini method with early stopping~\cite{bayes}.
In this method the regularisation parameter is the number of iterations used, which is optimised to obtain a relative $\chi^{2}$ variation between iterations lower than 5\%.

The ratio $R(x)$ of the single-diffractive to inclusive dijet cross sections is evaluated as a function of $x$ as:
\begin{equation}
 R(x)=\frac{\sigma^{\Pp\PX}_{\mathrm{jj}}(x)/\Delta\xi}{\sigma_{\mathrm{jj}}(x)} = \frac{ \mathcal{U}\left\{ N^{\Pp\PX}_{\mathrm{jj}}/A_{\text{\tiny CMS-TOTEM}} \right\}/\Delta\xi }{ \mathcal{U}\left\{ N_{\mathrm{jj}}/A_{\text{\tiny CMS}} \right\} },
 \label{eq:results:ratio-acceptance}
\end{equation}
where $N^{\Pp\PX}_{\mathrm{jj}}$ is the number of single-diffractive dijet candidates with $\xi_{\text{\tiny TOTEM}} < 0.1$, and $N_{\mathrm{jj}}$ is the total number of dijet events without the requirement of a proton detected in the RPs.
This number is dominated by the nondiffractive contribution.
The symbol $A_{\text{\tiny CMS-TOTEM}}$ indicates the acceptance of CMS and TOTEM for single-diffractive dijet events, evaluated with \POMWIG, \PYTHIAviii 4C and \PYTHIAviii CUETP8M1;
$A_{\text{\tiny CMS}}$ is the acceptance for nondiffractive dijet production ($\pt > 40\GeV$, $\abs{\eta} < 4.4$), evaluated with \PYTHIAvi, \PYTHIAviii 4C, \PYTHIAviii CUETP8M1, \PYTHIAviii CUETP8S1, and \HERWIGvi.
The acceptance includes unfolding corrections to the data with the D'Agostini method with early stopping, denoted by the symbol~$\mathcal{U}$ in Eq.~(\ref{eq:results:ratio-acceptance}).

\subsection{Systematic uncertainties}
\label{section:results-systematics}

The systematic uncertainties are estimated by varying the selections and modifying the analysis procedure, as discussed in this Section.
Tables~\ref{table:results_systematics_t_xi} and~\ref{table:ratio_systematics_x} summarise the main systematic uncertainties of the single-diffractive cross section and the ratio of the single-diffractive and inclusive dijet cross sections, respectively, presented in Sections~\ref{section:results-dsigmadtdxi} and~\ref{section:results-sd-nd-ratio}.

\begin{itemize}

 \item \textit{Trigger efficiency:}
 The trigger efficiency is calculated as a function of the subleading jet $\pt$ using a fit to the data.
 The sensitivity to the trigger efficiency determination is estimated by varying the fit parameters within their uncertainties.
 This variation corresponds to a trigger efficiency that increases or decreases by roughly 2\% at jet $\pt = 40\GeV$ and less than 1\% at $\pt = 50\GeV$.

 \item \textit{Calorimeter energy scale:}
 The reconstruction of $\xi_{\text{\tiny CMS}}$ is affected by the uncertainty in the calorimeter energy scale and is dominated by the HF contribution.
 This uncertainty is estimated by changing the energy of the PF candidates by $\pm 10\%$~\cite{Chatrchyan:2012vc,Chatrchyan:2011wm}.

 \item \textit{Jet energy scale and resolution:} The energy of the reconstructed jets is varied according to the jet energy scale uncertainty following the procedure described in Ref.~\cite{Khachatryan:2016kdb}.
 The systematic uncertainty in the jet energy resolution is estimated by varying the scale factors applied to the MC, as a function of pseudorapidity.
 The uncertainties obtained from the jet energy scale and resolution are added in quadrature.
 The effect of the jet energy resolution uncertainty amounts to less than 1\% of the measured cross section.

 \item \textit{Background:}
 Half the difference between the results of the ZB and HERA methods used to estimate the background, described in Section~\ref{section:background}, is an estimate of the effect of the systematic uncertainty of the background.

 \item \textit{RP acceptance:} The sensitivity to the size of the fiducial region for the impact position of the proton in the RPs is estimated by modifying its vertical boundaries by $200\mum$ and by reducing the horizontal requirement by 1\mm, to $0 < x < 6\mm$.
 Half the difference of the results thus obtained and the nominal ones is used as a systematic uncertainty. The uncertainties obtained when modifying the vertical and horizontal boundaries are added in quadrature.

 \item \textit{Resolution:} The reconstructed variables $t$ and $\xi$ are calculated by applying two methods: either directly, with a resolution function depending on each of these variables, or indirectly from the scattering angles $\theta_x^{\ast}$ and $\theta_y^{\ast}$. Half the difference between the results using the two methods is taken as a systematic uncertainty.

 \item \textit{Horizontal dispersion:}
 The reconstructed $\xi$ value depends on the optical functions describing the transport of the protons from the interaction vertex to the RP stations, and specifically the horizontal dispersion.
 This uncertainty is calculated by scaling the value of $\xi$ by $\pm 10\%$.
 This value corresponds to a conservative limit of the possible horizontal dispersion variation with respect to the nominal optics.

 \item \textit{$t$-slope:}
 The sensitivity to the modelling of the exponential $t$-slope is quantified by replacing its value in \POMWIG by that measured in the data.
 Half the difference between the results thus found and the nominal results is used as an estimate of the uncertainty.

 \item \textit{$\beta$-reweighting:}
 Half the difference of the results with and without the reweighting as a function of $\beta$ in \POMWIG (as discussed in Section~\ref{section:mc-simulation-acceptance}) is included in the systematic uncertainty.
 The effect amounts to less than 1\% of the single-diffractive cross section and less than about 6\% of the single-diffractive to inclusive dijet cross section ratio versus $x$.

 \item \textit{Acceptance and unfolding:}
 Half the maximum difference between the single-diffractive cross section results found by unfolding with \POMWIG, \PYTHIAviii 4C, and \PYTHIAviii CUETP8M1 is taken as a further component of the systematic uncertainty. Likewise for the results obtained with \PYTHIAvi Z2*, \PYTHIAviii 4C, \PYTHIAviii CUETP8M1 and \PYTHIAviii CUETP8S1 for the inclusive dijet cross section.

 \item \textit{Unfolding regularisation:}
 The regularisation parameter used in the unfolding, given by the number of iterations in the D'Agostini method used in this analysis,
 is optimised by calculating the relative $\chi^2$ variation between iterations.
 The value is chosen such that the $\chi^2$ variation is below 5\%.
 The number of iterations when the relative variation of $\chi^2$ is below 2\% is also used and half the difference with respect to the nominal is taken as a systematic uncertainty.

 \item \textit{Unfolding bias:}
 A simulated sample, including all detector effects, is unfolded with a different model.
 The difference between the corrected results thus obtained and those at the particle level is an estimate of the bias introduced by the unfolding procedure.
 Half the maximum difference obtained when repeating the procedure with all generator combinations is a measure of the systematic uncertainty related to the unfolding.

 \item \textit{Integrated luminosity:} The uncertainty in the integrated luminosity is 4\%, measured using a dedicated sample collected by TOTEM during the same data taking period~\cite{totem3}.

\end{itemize}

The total systematic uncertainty is calculated as the quadratic sum of the individual contributions.
The uncertainties in the jet energy scale and horizontal dispersion are the dominant contributions overall.

\clearPageSingle
\subsection{Extraction of the cross section as a function of \texorpdfstring{$t$ and $\xi$}{t and xi}}
\label{section:results-dsigmadtdxi}

\begin{figure*}[hbtp]
  \centering
    \includegraphics[width=\cmsFigWidth]{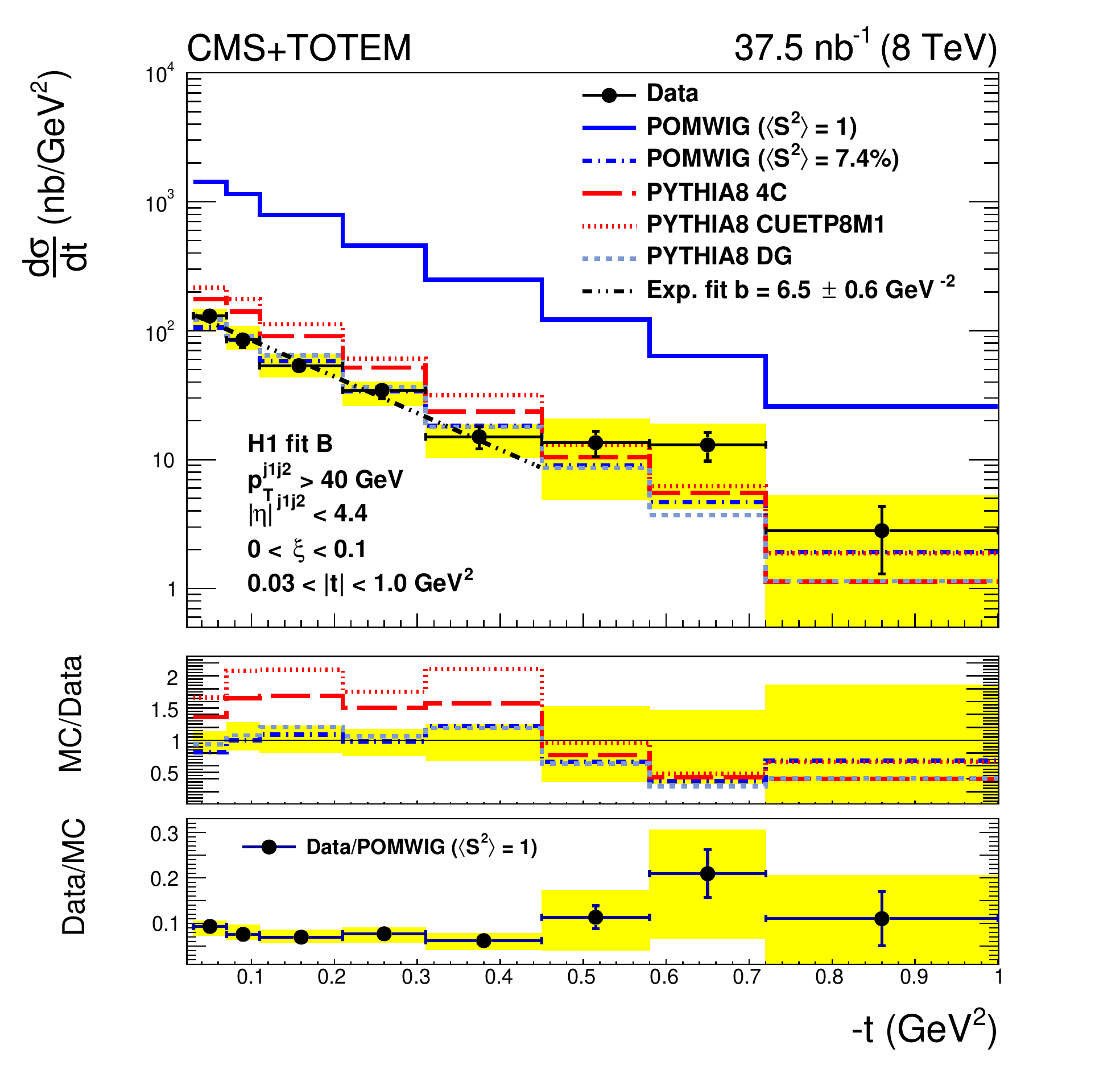}
    \includegraphics[width=\cmsFigWidth]{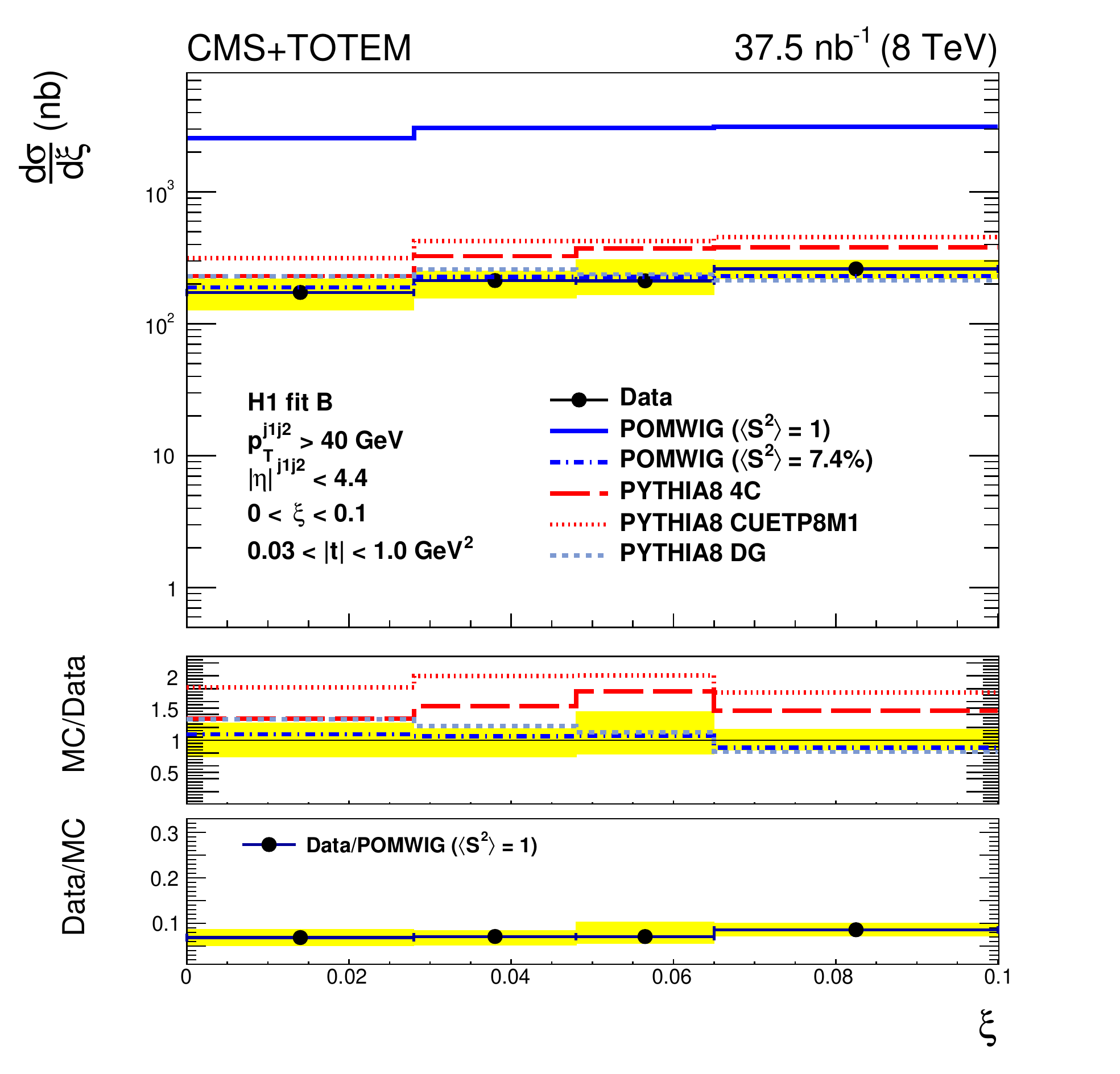}
     \caption{
     Differential cross section as a function of $t$ (left) and as a function of $\xi$ (right) for single-diffractive dijet production,
     compared to the predictions from \POMWIG, \PYTHIAviii 4C, \PYTHIAviii CUETP8M1, and \PYTHIAviii DG.
     The \POMWIG prediction is shown with no correction for the rapidity gap survival probability ($\langle S^{2} \rangle = 1$) and with a correction of $\langle S^{2} \rangle = 7.4\%$.
     The vertical bars indicate the statistical uncertainties and the yellow band indicates the total systematic uncertainty.
     The average of the results for events in which the proton is detected on either side of the interaction point is shown.
     The ratio between the data and the \POMWIG prediction, when no correction for the rapidity gap survival probability is applied, is shown in the bottom.
     }
    \label{fig:results:sigma_t_xi_result}
\end{figure*}

Figure~\ref{fig:results:sigma_t_xi_result} shows the differential cross section as a function of $t$ and $\xi$, integrated over the conjugate variable.
The results from events in which the proton is detected on either side of the IP are averaged.

The data are compared to \POMWIG, \PYTHIAviii 4C, \PYTHIAviii CUETP8M1, and \PYTHIAviii DG.
The \POMWIG prediction is shown for two values of the suppression of the diffractive cross section, \ie the rapidity gap survival probability, represented by $\langle S^{2} \rangle$.
When $\langle S^{2} \rangle = 1$, no correction is applied.
The resulting cross sections are higher than the data by roughly an order of magnitude, in agreement with the Tevatron results~\cite{diff_fermi3,diff_fermi4,ratio_cdf}.
The \POMWIG prediction is also shown with the correction $\langle S^{2} \rangle = 7.4\%$, calculated from the ratio of the measured diffractive cross section and the MC prediction, as discussed below.
After this correction, \POMWIG gives a good description of the data.
The \POMWIG prediction is shown in Fig.~\ref{fig:results:sigma_t_xi_result} as the sum of the Pomeron ($\Pp\Ppom$), Reggeon ($\Pp\Preg$) and Pomeron-Pomeron ($\Ppom\Ppom$) exchange contributions, while \PYTHIAviii includes only the Pomeron ($\Pp\Ppom$) contribution.
\PYTHIAviii 4C and \PYTHIAviii CUETP8M1 predict cross sections higher than the data by up to a factor of two.
The \PYTHIAviii DG model shows overall a good agreement with the data.
No correction is applied to the normalisation of the \PYTHIAviii samples.
The \PYTHIAviii DG model is the only calculation that predicts the cross section normalisation without an additional correction.

The ratio between the data and the \POMWIG predictions is shown in the bottom of the left and right panels of Fig.~\ref{fig:results:sigma_t_xi_result}.
No correction is applied for the rapidity gap survival probability ($\langle S^{2} \rangle = 1$).
Within the uncertainties, no significant dependence on $t$ and $\xi$ is observed.

The value of the cross section for single-diffractive dijet production, measured in the kinematic region $\pt > 40\GeV$, $\abs{\eta} < 4.4$, $\xi < 0.1$ and $0.03 < \abs{t} < 1\GeV^2$, is:
\begin{equation}
 \sigma^{\Pp\PX}_\mathrm{jj}=21.7 \pm 0.9\stat \,^{+3.0}_{-3.3}\syst \pm 0.9\lum \unit{nb}.
 \label{eq:results:single-diffractive-cross-section}
\end{equation}
Table~\ref{table:results_systematics_t_xi} summarises the main systematic uncertainties of the measured cross section.
The cross section is calculated independently for events in which the proton scatters towards the positive and negative $z$ directions, namely the processes $\Pp\Pp \to \Pp\PX$ and $\Pp\Pp \to \PX\Pp$, and the results are averaged. They are compatible within the uncertainties.
The \PYTHIAviii DG model predicts in the same kinematic region a cross section of $23.7\unit{nb}$, consistent with the measurement.

\begin{table}[hbtp]
 \centering
  \topcaption{
  Individual contributions to the systematic uncertainties in the measurement of the single-diffractive dijet production cross section in the kinematic region $\pt > 40\GeV$, $\abs{\eta} < 4.4$, $\xi < 0.1$, and $0.03 < \abs{t} < 1\GeV^2$.
  The second column indicates the relative uncertainties in the integrated cross section. The third and fourth columns represent the minimum and maximum relative uncertainties in the differential cross sections in bins of $t$ and $\xi$, respectively.
  The minimum relative uncertainty is not shown when it is below 1\%.
  The total uncertainty is the quadratic sum of the individual contributions.
  The uncertainty of the integrated luminosity is not shown.
  }
  \label{table:results_systematics_t_xi}
  \cmsTable{
  \begin{tabular}{lccc}
   \hline
   \multirow{2}{*}{Uncertainty source}    & \multicolumn{3}{c}{Relative uncertainty} \\
                                          & $\sigma^{\Pp\PX}_\mathrm{jj}$ & $\rd\sigma/\rd{}t$ & $\rd\sigma/\rd\xi$ \\
   \hline
   Trigger efficiency        		  & $\pm$2 \%       & 1--2\%       & $<$2.4\%     \\
   Calorimeter energy scale  		  & $+$1/$-$2 \%    & $<$7\%       & $<$7\%       \\
   Jet energy scale and resolution        & $+$9/$-$8 \%    & 3--32\%      & 7--16\%      \\
   Background       		          & $\pm$3 \%       & 2--27\%      & $<$8\%       \\
   RP acceptance      		          & $<$1 \%         & $<$21\%      & $<$2\%       \\
   Resolution        			  & $\pm$2 \%       & 2--30\%      & $<$8\%       \\
   Horizontal dispersion   		  & $+$9/$-$12 \%   & 8--71\%      & 8--41\%      \\
   $t$-slope               		  & $<$1 \%         & $<$16\%      & $<$1.3\%     \\
   $\beta$-reweighting    		  & $<$1 \%         & $<$1\%       & $<$1\%       \\
   Acceptance and unfolding  		  & $\pm$2 \%       & 2--50\%      & 5--12\%      \\
   Unfolding bias           		  & $\pm$3 \%       & 2--50\%      & 5--11\%      \\
   Unfolding regularization               & \NA             & $<$8\%       & $<$1\%       \\
   Total                   		  & $+$14/$-$15 \%  &              &              \\
   \hline
  \end{tabular}
  }
\end{table}

The differential cross section as a function of $t$ is well described by an exponential function for $\abs{t}$ values up to about $0.4\GeV^2$.
A fit is performed with the function $\rd\sigma/{\rd}t \propto \exp\left({-b\abs{t}}\right)$ for $t$ values in the range $0.03 < \abs{t} < 0.45\GeV^2$.

The resulting exponential slope is:
\begin{equation}
 b = 6.5 \pm 0.6\stat \,^{+1.0}_{-0.8}\syst \GeV^{-2},
 \label{eq:results:t-slope}
\end{equation}
where the systematic uncertainties include the contributions discussed in Section~\ref{section:results-systematics}.
The results for the exponential slope of the cross section calculated independently for events in which the proton scatters towards the positive and negative $z$ directions are compatible within the uncertainties.

The parametrisation obtained from the fit is shown in Fig.~\ref{fig:results:sigma_t_xi_result}. In the fit range ($0.03 < \abs{t} < 0.45\GeV^2$), the horizontal position of the data points is calculated as the value for which the parametrised function equals its average over the bin width.
The data points in the larger-$\abs{t}$ region outside the fit range ($\abs{t} > 0.45\GeV^2$) are shown at the centre of the bins.

The slope measured by CDF is $b \approx 5\text{--}6\GeV^{-2}$ for $\abs{t} \lessapprox 0.5\GeV^2$~\cite{diff_fermi4}.
In the larger-$\abs{t}$ region, the CDF data exhibit a smaller slope that becomes approximately independent of $t$ for $\abs{t} \gtrapprox 2\GeV^2$.

The present measurement of the slope is consistent with that by CDF at small-$\abs{t}$.
The data do not conclusively indicate a flattening of the $t$ distribution at larger-$\abs{t}$.

An estimate of the rapidity gap survival probability can be obtained from the ratio of the measured cross section in Eq.~(\ref{eq:results:single-diffractive-cross-section}) and that predicted by \POMWIG with $\langle S^{2} \rangle = 1$.  Alternatively, the \PYTHIAviii hard-diffraction model can be used if the DG suppression framework is not applied.
The two results are consistent.

The overall suppression factor obtained with respect to the \POMWIG cross section is $\langle S^{2} \rangle = 7.4 \,^{+1.0}_{-1.1}\%$, where the statistical and systematic uncertainties are added in quadrature.
A similar result is obtained when the \PYTHIAviii unsuppressed cross section is used as reference value.

The H1 fit B dPDFs used in this analysis include the contribution from proton dissociation in $\Pe\Pp$ collisions. They are extracted from the process $\Pe\Pp \to \Pe\PX\mathrm{Y}$, where Y can be a proton or a low-mass excitation with $M_{\mathrm{Y}} < 1.6\GeV$~\cite{diff_desy1}.
The results found when the proton is detected are consistent, apart from a different overall normalisation. The ratio of the cross sections is $\sigma( M_{\mathrm{Y}} < 1.6\GeV )/\sigma( M_{\mathrm{Y}} = M_{\Pp} ) = 1.23 \pm 0.03\stat \pm 0.16\syst$~\cite{diff_desy1,Aktas:2006hx}. No dependence on $\beta$, $Q^2$, or $\xi$ is observed.
To account for the different normalisation, the ratio is used to correct $\langle S^{2} \rangle$; this yields $\langle S^{2} \rangle = \left( 9 \pm 2 \right)\%$ when the \POMWIG cross section is taken as the reference value.
A similar result is obtained with \PYTHIAviii.

\subsection{Extraction of the ratio of the single-diffractive to inclusive dijet yields}
\label{section:results-sd-nd-ratio}

\begin{figure*}[hbtp]
  \centering
    \includegraphics[width=\cmsFigWidth]{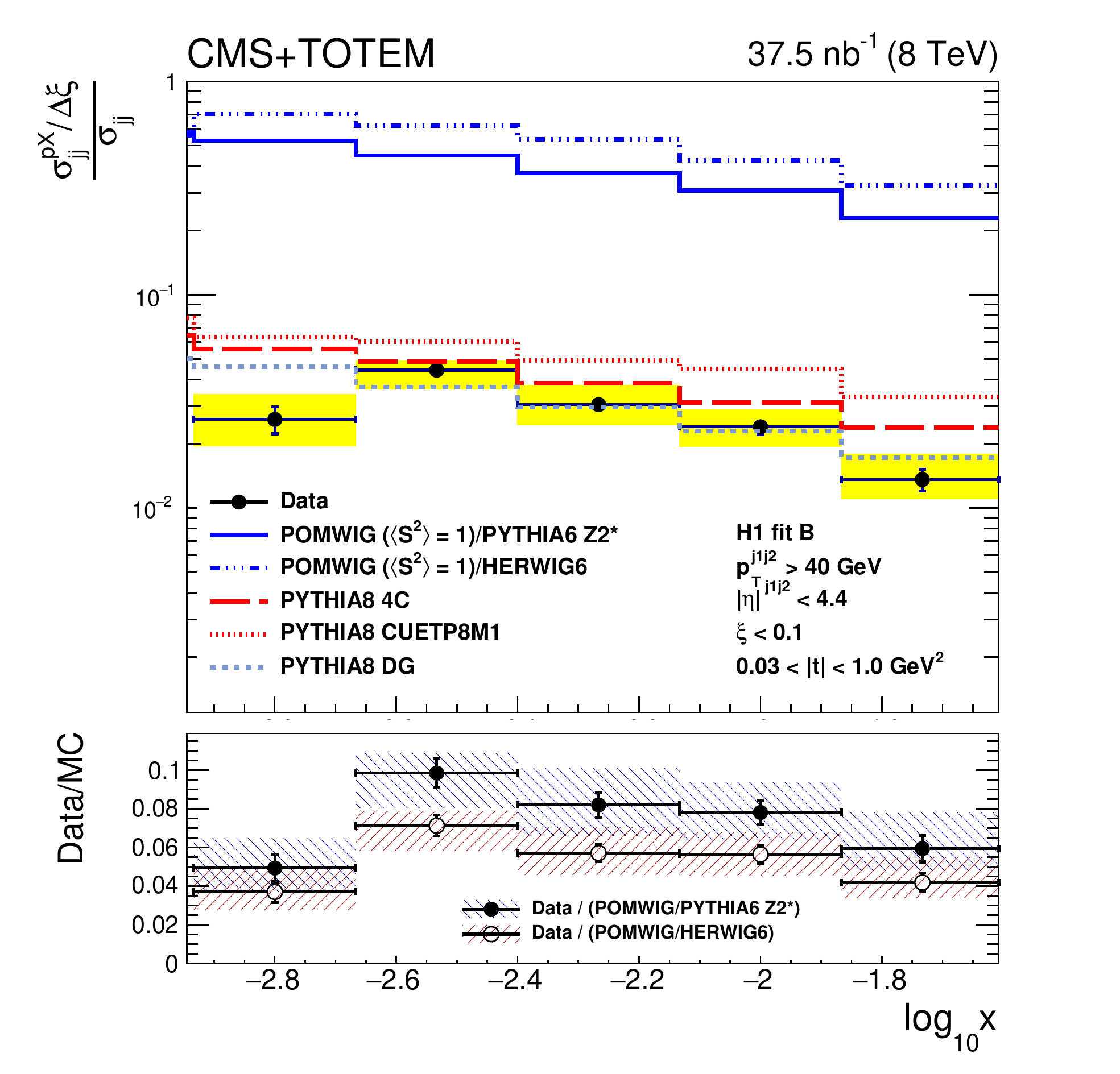}
    \includegraphics[width=\cmsFigWidth]{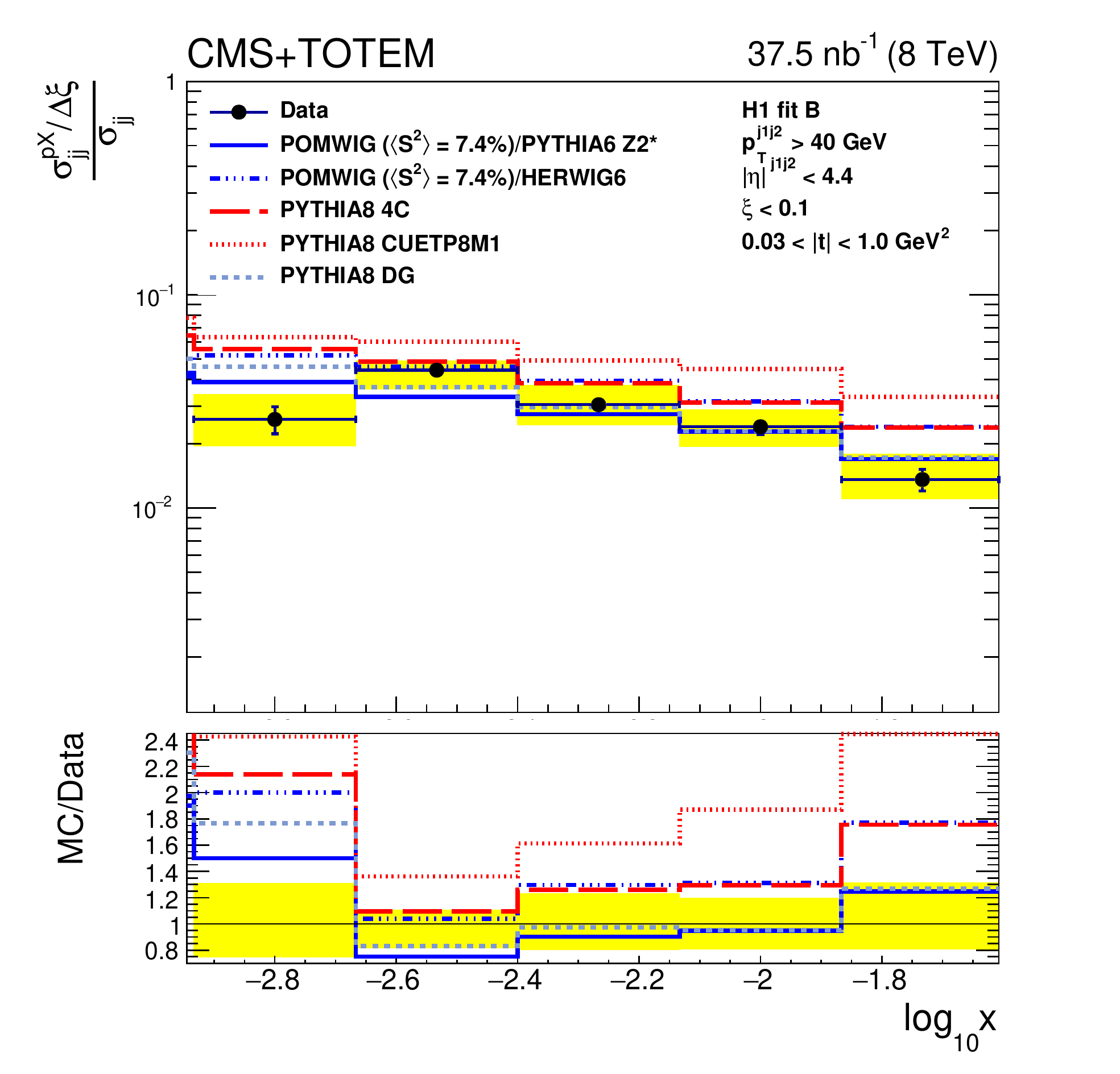}
     \caption{
     Ratio per unit of $\xi$ of the single-diffractive and inclusive dijet cross sections in the region given by $\xi < 0.1$ and $0.03 < \abs{t} < 1\GeV^2$,
     compared to the predictions from the different models for the ratio between the single-diffractive and nondiffractive cross sections.
     The \POMWIG prediction is shown with no correction for the rapidity gap survival probability ($\langle S^{2} \rangle = 1$) (left) and with a correction of $\langle S^{2} \rangle = 7.4\%$ (right).
     The vertical bars indicate the statistical uncertainties and the yellow band indicates the total systematic uncertainty.
     The average of the results for events in which the proton is detected on either side of the interaction point is shown.
     The ratio between the data and the \POMWIG prediction using \PYTHIAvi or \HERWIGvi as the nondiffractive contribution, when no correction for the rapidity gap survival probability is applied, is shown in the bottom of the left panel.
     }
    \label{fig:results:ratio_average}
\end{figure*}

Figure~\ref{fig:results:ratio_average} shows the ratio $R(x)$ in the kinematic region $\pt > 40\GeV$, $\abs{\eta} < 4.4$, $\xi < 0.1$, $0.03 < \abs{t} < 1\GeV^2$ and $-2.9\leq\log_{10} x\leq-1.6$.
The average of the results for events in which the proton is detected on either side of the IP is shown.
The yellow band represents the total systematic uncertainty (\cf Section~\ref{section:results-systematics}).
The data are compared to the ratio of the single-diffractive and nondiffractive dijet cross sections from different models.
The single-diffractive contribution is simulated with \POMWIG, \PYTHIAviii 4C, \PYTHIAviii CUETP8M1, and \PYTHIAviii DG.
The nondiffractive contribution is simulated with \PYTHIAvi and \HERWIGvi if \POMWIG is used for the diffractive contribution.
When using \PYTHIAviii the diffractive and nondiffractive contributions are simulated with the same underlying event tune.
When no correction for the rapidity gap survival probability is applied ($\langle S^{2} \rangle = 1$), \POMWIG gives a ratio higher by roughly an order of magnitude, consistent with the results discussed in Section~\ref{section:results-dsigmadtdxi}.
The suppression seen in the data with respect to the simulation is not substantially different when using \PYTHIAvi or \HERWIGvi for the nondiffractive contribution.
\POMWIG with a correction of $\langle S^{2} \rangle = 7.4\%$ gives overall a good description of the data when \PYTHIAvi is used for the nondiffractive contribution.
When \HERWIGvi is used for the nondiffractive contribution the agreement is worse, especially in the lower- and higher-$x$ regions.
The agreement for \PYTHIAviii 4C is fair in the intermediate $x$ region, but worse at low- and high-$x$.
The agreement is worse for \PYTHIAviii CUETP8M1, with values of the ratio higher than those in the data by up to a factor of two.
The \PYTHIAviii DG predictions agree well with the data overall, though the agreement is worse in the lowest-$x$ bin.
No correction is applied to the \PYTHIAviii normalisation.
In the lowest-$x$ bin, the ratio in the data is below the predictions. 
The observed discrepancy is not significant for the predictions that agree well overall with the data elsewhere, taking into account the systematic and statistical uncertainties.

The measured value of the ratio, normalised per unit of $\xi$, in the full kinematic region defined above is:
\begin{equation}
R = \left(\sigma^{\Pp\PX}_{\mathrm{jj}}/\Delta\xi\right)/\sigma_{\mathrm{jj}} = 0.025 \pm 0.001\stat \pm 0.003\syst.
\end{equation}
Table~\ref{table:ratio_systematics_x} summarises the main contributions to the systematic uncertainty of the ratio.
The uncertainty of the jet energy scale is considerably smaller than in the case of the single-diffractive cross section.

\begin{table}[hbtp]
 \centering
  \topcaption{
   Individual contributions to the systematic uncertainty in the measurement of the single-diffractive to inclusive dijet yields ratio in the kinematic region $\pt > 40\GeV$, $\abs{\eta} < 4.4$, $\xi < 0.1$, $0.03 < \abs{t} < 1\GeV^2$, and $-2.9 \leq \log_{10} x \leq -1.6$.
   The second and third columns represent the relative uncertainties in the ratio in the full kinematic region and in bins of $\log_{10}x$, respectively. 
   The minimum relative uncertainty is not shown when it is below 1\%.
   The total uncertainty is the quadratic sum of the individual contributions.
   }
  \label{table:ratio_systematics_x}
  \begin{tabular}{lcc}
   \hline
   \multirow{2}{*}{Uncertainty source}    & \multicolumn{2}{c}{Relative uncertainty} \\
                                          & $R$ & $R(x)$ \\
   \hline
   Trigger efficiency   		  & Negligible       &  2--3\%            \\
   Calorimeter energy scale 		  & $+$1/$-$2 \%     &  $<$7\%            \\
   Jet energy scale and resolution        & $\pm$2 \%        &  1--10\%           \\
   Background       		          & $\pm$1 \%        &  1--17\%           \\
   RP acceptance      			  & $<$1 \%          &  $<$4\%            \\
   Resolution             		  & $\pm$2 \%        &  $<$4\%            \\
   Horizontal dispersion   		  & $+$9/$-$11 \%    &  11--23\%          \\
   $t$-slope              		  & $<$1 \%          &  $<$3\%            \\
   $\beta$-reweighting     		  & $\pm$1 \%        &  $<$6\%            \\
   Acceptance and unfolding 		  & $\pm$2 \%        &  3--11\%           \\
   Unfolding bias        		  & $\pm$3 \%        &  3--14\%           \\
   Unfolding regularization               & \NA              &  $<$11\%           \\
   Total            			  & $+$10/$-$13 \%   &                    \\
   \hline
  \end{tabular}
\end{table}

\begin{figure}[hbtp]
  \centering
    \includegraphics[width=\cmsFigWidthSingleLarge]{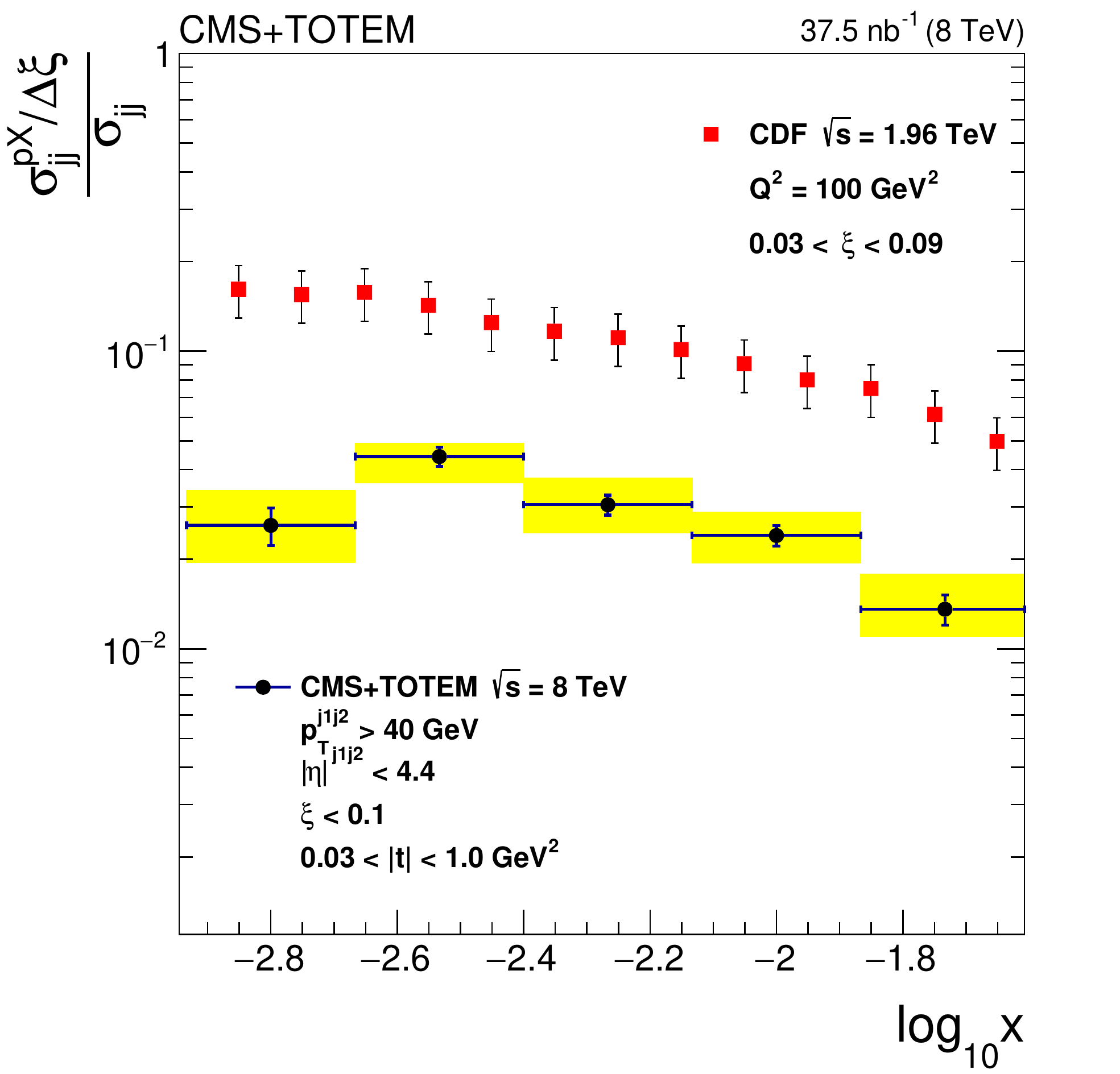}
     \caption{
      Ratio per unit of $\xi$ of the single-diffractive and inclusive dijet cross sections in the kinematic region given by $\xi < 0.1$ and $0.03 < \abs{t} < 1\GeV^2$.
      The vertical bars indicate the statistical uncertainties and the yellow band indicates the total systematic uncertainty.
      The red squares represent the results obtained by CDF at $\sqrt{s} = 1.96\TeV$ for jets with $Q^2 \approx 100\GeV^2$ and $\abs{\eta} < 2.5$, with $0.03 < \xi < 0.09$.
     }
    \label{fig:ratio_cdf}
\end{figure}

Figure~\ref{fig:ratio_cdf} shows the comparison between the results of Fig.~\ref{fig:results:ratio_average} and those from CDF~\cite{diff_fermi4}.
The CDF results are shown for jets with $Q^2$ of roughly $100\GeV^2$ and pseudorapidity $\abs{\eta} < 2.5$, with $0.03 < \xi < 0.09$. In this case $Q^2$ is defined, per event, as the mean transverse energy of the two leading jets squared. CDF measures the ratio for $Q^2$ values up to $10^4\GeV^2$. A relatively small dependence on $Q^2$ is observed.
The present data are lower than the CDF results.
A decrease of the ratio of diffractive to inclusive cross sections with centre-of-mass energy has also been observed by CDF by comparing data at 630 and 1800\GeV~\cite{ratio_cdf}.

\section{Summary}

The differential cross section for single-diffractive dijet production in proton-proton ($\Pp\Pp$) collisions at $\sqrt{s} = 8\TeV$ has been measured as a function of the proton fractional momentum loss $\xi$ and the squared four momentum transfer $t$, using the CMS and TOTEM detectors.
The data, corresponding to an integrated luminosity of $37.5\nbinv$, were collected using a nonstandard optics configuration with $\beta^* = 90\unit{m}$.
The processes considered are $\Pp\Pp \to \Pp\PX$ or $\Pp\Pp \to \PX\Pp$, with $\PX$ including a system of two jets, in the kinematic region $\xi<0.1$ and $0.03 < \abs{t} < 1.0\GeV^2$.
The two jets have transverse momentum $\pt > 40\GeV$ and pseudorapidity $\abs{\eta} < 4.4$.
The integrated cross section in this kinematic region is
$\sigma^{\Pp\PX}_\mathrm{jj} = 21.7 \pm 0.9\stat \,^{+3.0}_{-3.3}\syst \pm 0.9\lum \unit{nb}$;
it is the average of the cross sections when the proton scatters to either side of the interaction point.
The exponential slope of the cross section as a function of $t$ is
$b = 6.5 \pm 0.6\stat \,^{+1.0}_{-0.8}\syst \GeV^{-2}$.
This is the first measurement of hard diffraction with a measured proton at the LHC.

The data are compared with the predictions of different models. After applying a normalisation shift ascribed to the rapidity gap survival probability, \POMWIG agrees well with the data.
The \PYTHIAviii dynamic gap model describes the data well, both in shape and normalisation.
In this model the effects of the rapidity gap survival probability are simulated within the framework of multiparton interactions.
The \PYTHIAviii dynamic gap model is the only calculation that predicts the cross section normalisation without an additional correction.

The ratios of the measured single-diffractive cross section to those predicted by \POMWIG and \PYTHIAviii give estimates of the rapidity gap survival probability.
After accounting for the correction of the dPDF normalisation due to proton dissociation, the value of $\langle S^{2} \rangle$ is $\left( 9 \pm 2 \right)\%$ when using \POMWIG as the reference cross section value, with a similar result when \PYTHIAviii is used.

The ratio of the single-diffractive to inclusive dijet cross section has been measured as a function of the parton momentum fraction $x$.
The ratio is lower than that observed at CDF at a smaller centre-of-mass energy.
In the region $\pt > 40\GeV$, $\abs{\eta} < 4.4$, $\xi < 0.1$, $0.03 < \abs{t} < 1.0\GeV^2$, and $-2.9 \leq \log_{10} x \leq -1.6$, the ratio, normalised per unit $\xi$, is
$R = (\sigma^{\Pp\PX}_{\mathrm{jj}}/\Delta\xi)/\sigma_{\mathrm{jj}} = 0.025 \pm 0.001\stat \pm 0.003\syst$.

\begin{acknowledgments}
We congratulate our colleagues in the CERN accelerator departments for the excellent performance of the LHC and thank the technical and administrative staffs at CERN and at other CMS and TOTEM institutes for their contributions to the success of the common CMS-TOTEM effort. We gratefully acknowledge work of the beam optics development team at CERN for the design and the successful commissioning of the high $\beta^*$ optics and thank the LHC machine coordinators for scheduling dedicated fills. In addition, we gratefully acknowledge the computing centres and personnel of the Worldwide LHC Computing Grid for delivering so effectively the computing infrastructure essential to our analyses. Finally, we acknowledge the enduring support for the construction and operation of the LHC and the CMS and TOTEM detectors provided by the following funding agencies: BMBWF and FWF (Austria); FNRS and FWO (Belgium); CNPq, CAPES, FAPERJ, FAPERGS, and FAPESP (Brazil); MES (Bulgaria); CERN; CAS, MoST, and NSFC (China); COLCIENCIAS (Colombia); MSES and CSF (Croatia); RPF (Cyprus); SENESCYT (Ecuador); MoER, ERC IUT, PUT and ERDF (Estonia); Academy of Finland, Finnish Academy of Science and Letters (The Vilho Yrj{\"o} and Kalle V{\"a}is{\"a}l{\"a} Fund), MEC, Magnus Ehrnrooth Foundation, HIP, and Waldemar von Frenckell Foundation (Finland); CEA and CNRS/IN2P3 (France); BMBF, DFG, and HGF (Germany); GSRT (Greece); the Circles of Knowledge Club, NKFIA (Hungary); DAE and DST (India); IPM (Iran); SFI (Ireland); INFN (Italy); MSIP and NRF (Republic of Korea); MES (Latvia); LAS (Lithuania); MOE and UM (Malaysia); BUAP, CINVESTAV, CONACYT, LNS, SEP, and UASLP-FAI (Mexico); MOS (Montenegro); MBIE (New Zealand); PAEC (Pakistan); MSHE and NSC (Poland); FCT (Portugal); JINR (Dubna); MON, RosAtom, RAS, RFBR, and NRC KI (Russia); MESTD (Serbia); SEIDI, CPAN, PCTI, and FEDER (Spain); MOSTR (Sri Lanka); Swiss Funding Agencies (Switzerland); MST (Taipei); ThEPCenter, IPST, STAR, and NSTDA (Thailand); TUBITAK and TAEK (Turkey); NASU (Ukraine); STFC (United Kingdom); DOE and NSF (USA).

\hyphenation{Rachada-pisek} Individuals have received support from the Marie-Curie programme and the European Research Council and Horizon 2020 Grant, contract Nos.\ 675440, 752730, and 765710 (European Union); the Leventis Foundation; the A.P.\ Sloan Foundation; the Alexander von Humboldt Foundation; the Belgian Federal Science Policy Office; the Fonds pour la Formation \`a la Recherche dans l'Industrie et dans l'Agriculture (FRIA-Belgium); the Agentschap voor Innovatie door Wetenschap en Technologie (IWT-Belgium); the F.R.S.-FNRS and FWO (Belgium) under the ``Excellence of Science -- EOS" -- be.h project n.\ 30820817; the Beijing Municipal Science \& Technology Commission, No. Z191100007219010; the Ministry of Education, Youth and Sports (MEYS) and MSMT CR of the Czech Republic; the Nylands nation vid Helsingfors universitet (Finland); the Deutsche Forschungsgemeinschaft (DFG) under Germany’s Excellence Strategy -- EXC 2121 ``Quantum Universe" -- 390833306; the Lend\"ulet (``Momentum") Programme and the J\'anos Bolyai Research Scholarship of the Hungarian Academy of Sciences, the New National Excellence Program \'UNKP, the NKFIA research grants 123842, 123959, 124845, 124850, 125105, 128713, 128786, 129058, K 133046, and EFOP-3.6.1- 16-2016-00001 (Hungary); the Council of Science and Industrial Research, India; the HOMING PLUS programme of the Foundation for Polish Science, cofinanced from European Union, Regional Development Fund, the Mobility Plus programme of the Ministry of Science and Higher Education, including Grant No. MNiSW DIR/WK/2018/13, the National Science Center (Poland), contracts Harmonia 2014/14/M/ST2/00428, Opus 2014/13/B/ST2/02543, 2014/15/B/ST2/03998, and 2015/19/B/ST2/02861, Sonata-bis 2012/07/E/ST2/01406; the National Priorities Research Program by Qatar National Research Fund; the Ministry of Science and Education, grant no. 14.W03.31.0026 (Russia); the Programa Estatal de Fomento de la Investigaci{\'o}n Cient{\'i}fica y T{\'e}cnica de Excelencia Mar\'{\i}a de Maeztu, grant MDM-2015-0509 and the Programa Severo Ochoa del Principado de Asturias; the Thalis and Aristeia programmes cofinanced by EU-ESF and the Greek NSRF; the Rachadapisek Sompot Fund for Postdoctoral Fellowship, Chulalongkorn University and the Chulalongkorn Academic into Its 2nd Century Project Advancement Project (Thailand); the Kavli Foundation; the Nvidia Corporation; the SuperMicro Corporation; the Welch Foundation, contract C-1845; and the Weston Havens Foundation (USA).
\end{acknowledgments}

\bibliography{auto_generated}   

\providecommand{\href}[2]{#2}\begingroup\raggedright\begin{thebibliography}{10}%
\makeatletter
\providecommand{\hrefCMSnoop }[0]{\@secondoftwo}%
\makeatother
\providecommand{\doi}{\texttt{doi:}\begingroup \urlstyle{tt}\Url}

\bibitem{Collins:1977jy}
{P. D. B. Collins}, ``An Introduction to Regge Theory and High-Energy
  Physics''.
\newblock Cambridge Monographs on Mathematical Physics. Cambridge Univ. Press,
  Cambridge, UK, 2009.
\newblock ISBN~9780521110358.

\bibitem{diff_desy1}
\hrefCMSnoop {}{{H1} Collaboration, ``Measurement and {QCD} analysis of the
  diffractive deep-inelastic scattering cross section at {HERA}'',} \textit{
  Eur. Phys. J. C} \textbf{ 48} (2006) 715,
  \href{http://dx.doi.org/10.1140/epjc/s10052-006-0035-3}{\doi{10.1140/epjc/s10052-006-0035-3}},
  \href{http://www.arXiv.org/abs/hep-ex/0606004}{\texttt{arXiv:hep-ex/0606004}}.

\bibitem{Aktas:2006hx}
\hrefCMSnoop {}{{H1} Collaboration, ``Diffractive deep-inelastic scattering
  with a leading proton at {HERA}'',} \textit{ Eur. Phys. J. C} \textbf{ 48}
  (2006) 749,
  \href{http://dx.doi.org/10.1140/epjc/s10052-006-0046-0}{\doi{10.1140/epjc/s10052-006-0046-0}},
\href{http://www.arXiv.org/abs/hep-ex/0606003}{\texttt{arXiv:hep-ex/0606003}}.

\bibitem{diff_desy2}
\hrefCMSnoop {}{{ZEUS} Collaboration, ``Deep inelastic scattering with leading
  protons or large rapidity gaps at {HERA}'',} \textit{ Nucl. Phys. B} \textbf{
  816} (2009) 1,
  \href{http://dx.doi.org/10.1016/j.nuclphysb.2009.03.003}{\doi{10.1016/j.nuclphysb.2009.03.003}},
  \href{http://www.arXiv.org/abs/0812.2003}{\texttt{arXiv:0812.2003}}.

\bibitem{diff_desy3}
\hrefCMSnoop {}{{ZEUS} Collaboration, ``A {QCD} analysis of {ZEUS} diffractive
  data'',} \textit{ Nucl. Phys. B} \textbf{ 831} (2010) 1,
  \href{http://dx.doi.org/10.1016/j.nuclphysb.2010.01.014}{\doi{10.1016/j.nuclphysb.2010.01.014}},
  \href{http://www.arXiv.org/abs/0911.4119}{\texttt{arXiv:0911.4119}}.

\bibitem{diff_cern}
\hrefCMSnoop {}{{UA8} Collaboration, ``Evidence for a super-hard pomeron
  structure'',} \textit{ Phys. Lett. B} \textbf{ 297} (1992) 417,
\href{http://dx.doi.org/10.1016/0370-2693(92)91281-D}{\doi{10.1016/0370-2693(92)91281-D}}.

\bibitem{diff_fermi1}
\hrefCMSnoop {}{{CDF} Collaboration, ``Measurement of diffractive dijet
  production at the {Fermilab} {Tevatron}'',} \textit{ Phys. Rev. Lett.}
  \textbf{ 79} (1997) 2636,
  \href{http://dx.doi.org/10.1103/PhysRevLett.79.2636}{\doi{10.1103/PhysRevLett.79.2636}}.

\bibitem{diff_fermi2}
\hrefCMSnoop {}{{D0} Collaboration, ``Hard single diffraction in {$\PAp\Pp$}
  collisions at $\sqrt{s}$ = 630 and 1800 {GeV}'',} \textit{ Phys. Lett. B}
  \textbf{ 531} (2002) 52,
  \href{http://dx.doi.org/10.1016/S0370-2693(02)01364-3}{\doi{10.1016/S0370-2693(02)01364-3}},
  \href{http://www.arXiv.org/abs/hep-ex/9912061}{\texttt{arXiv:hep-ex/9912061}}.

\bibitem{diff_fermi3}
\hrefCMSnoop {}{{CDF} Collaboration, ``Diffractive dijets with a leading
  antiproton in {$\PAp\Pp$} collisions at $\sqrt{s}$ = 1800 {GeV}'',} \textit{
  Phys. Rev. Lett.} \textbf{ 84} (2000) 5043,
  \href{http://dx.doi.org/10.1103/PhysRevLett.84.5043}{\doi{10.1103/PhysRevLett.84.5043}}.

\bibitem{diff_fermi4}
\hrefCMSnoop {}{{CDF} Collaboration, ``Diffractive dijet production in
  {$\PAp\Pp$} collisions at $\sqrt{s}$ = 1.96 {TeV}'',} \textit{ Phys. Rev. D}
  \textbf{ 86} (2012) 032009,
  \href{http://dx.doi.org/10.1103/PhysRevD.86.032009}{\doi{10.1103/PhysRevD.86.032009}},
  \href{http://www.arXiv.org/abs/1206.3955}{\texttt{arXiv:1206.3955}}.

\bibitem{ratio_cdf}
\hrefCMSnoop {}{{CDF} Collaboration, ``Diffractive dijet production at
  $\sqrt{s}$ = 630 and $\sqrt{s}$ = 1800 {GeV} at the {Fermilab} {Tevatron}'',}
  \textit{ Phys. Rev. Lett.} \textbf{ 88} (2002) 151802,
  \href{http://dx.doi.org/10.1103/PhysRevLett.88.151802}{\doi{10.1103/PhysRevLett.88.151802}},
  \href{http://www.arXiv.org/abs/hep-ex/0109025}{\texttt{arXiv:hep-ex/0109025}}.

\bibitem{Chatrchyan:2012vc}
\hrefCMSnoop {}{{CMS Collaboration}, ``Observation of a diffractive
  contribution to dijet production in proton-proton collisions at $\sqrt{s}=7$
  {TeV}'',} \textit{ Phys. Rev. D} \textbf{ 87} (2013) 012006,
  \href{http://dx.doi.org/10.1103/PhysRevD.87.012006}{\doi{10.1103/PhysRevD.87.012006}},
\href{http://www.arXiv.org/abs/1209.1805}{\texttt{arXiv:1209.1805}}.

\bibitem{Aad:2015xis}
\hrefCMSnoop {}{{ATLAS Collaboration}, ``Dijet production in $\sqrt{s} = 7$
  {TeV} {$\Pp\Pp$} collisions with large rapidity gaps at the {ATLAS}
  experiment'',} \textit{ Phys. Lett. B} \textbf{ 754} (2016) 214,
  \href{http://dx.doi.org/10.1016/j.physletb.2016.01.028}{\doi{10.1016/j.physletb.2016.01.028}},
\href{http://www.arXiv.org/abs/1511.00502}{\texttt{arXiv:1511.00502}}.

\bibitem{diff_desy4}
\hrefCMSnoop {}{{H1} Collaboration, ``Diffractive dijet photoproduction in
  {$\Pe\Pp$} collisions at {HERA}'',} \textit{ Eur. Phys. J. C} \textbf{ 70}
  (2010) 15,
  \href{http://dx.doi.org/10.1140/epjc/s10052-010-1448-6}{\doi{10.1140/epjc/s10052-010-1448-6}},
  \href{http://www.arXiv.org/abs/1006.0946}{\texttt{arXiv:1006.0946}}.

\bibitem{dpdf1}
\hrefCMSnoop {}{{L. Trentadue and G. Veneziano}, ``Fracture functions. {An}
  improved description of inclusive hard processes in {QCD}'',} \textit{ Phys.
  Lett. B} \textbf{ 323} (1994) 201,
  \href{http://dx.doi.org/10.1016/0370-2693(94)90292-5}{\doi{10.1016/0370-2693(94)90292-5}}.

\bibitem{dpdf2}
\hrefCMSnoop {}{{J. C. Collins}, ``Proof of factorization for diffractive hard
  scattering'',} \textit{ Phys. Rev. D} \textbf{ 57} (1998) 3051,
  \href{http://dx.doi.org/10.1103/PhysRevD.57.3051}{\doi{10.1103/PhysRevD.57.3051}},
  \href{http://www.arXiv.org/abs/hep-ph/9709499}{\texttt{arXiv:hep-ph/9709499}}.
  [Erratum: \DOI{10.1103/PhysRevD.61.019902}].

\bibitem{dpdf3}
\hrefCMSnoop {}{{M. Grazzini, L. Trentadue, and G. Veneziano}, ``Fracture
  functions from cut vertices'',} \textit{ Nucl. Phys. B} \textbf{ 519} (1998)
  394,
  \href{http://dx.doi.org/10.1016/S0550-3213(97)00840-7}{\doi{10.1016/S0550-3213(97)00840-7}},
  \href{http://www.arXiv.org/abs/hep-ph/9709452}{\texttt{arXiv:hep-ph/9709452}}.

\bibitem{Bjorken:1992er}
\hrefCMSnoop {}{J.~D. Bjorken, ``Rapidity gaps and jets as a new-physics
  signature in very high-energy hadron-hadron collisions'',} \textit{ Phys.
  Rev. D} \textbf{ 47} (1993) 101,
\href{http://dx.doi.org/10.1103/PhysRevD.47.101}{\doi{10.1103/PhysRevD.47.101}}.

\bibitem{Sirunyan:2017ulk}
\hrefCMSnoop {}{{CMS Collaboration}, ``Particle-flow reconstruction and global
  event description with the {CMS} detector'',} \textit{ JINST} \textbf{ 12}
  (2017) P10003,
  \href{http://dx.doi.org/10.1088/1748-0221/12/10/P10003}{\doi{10.1088/1748-0221/12/10/P10003}},
\href{http://www.arXiv.org/abs/1706.04965}{\texttt{arXiv:1706.04965}}.

\bibitem{Cacciari:2008gp}
\hrefCMSnoop {}{M.~Cacciari, G.~P. Salam, and G.~Soyez, ``The anti-\kt jet
  clustering algorithm'',} \textit{ JHEP} \textbf{ 04} (2008) 063,
  \href{http://dx.doi.org/10.1088/1126-6708/2008/04/063}{\doi{10.1088/1126-6708/2008/04/063}},
  \href{http://www.arXiv.org/abs/0802.1189}{\texttt{arXiv:0802.1189}}.

\bibitem{Cacciari:2011ma}
\hrefCMSnoop {}{M.~Cacciari, G.~P. Salam, and G.~Soyez, ``{FastJet} user
  manual'',} \textit{ Eur. Phys. J. C} \textbf{ 72} (2012) 1896,
  \href{http://dx.doi.org/10.1140/epjc/s10052-012-1896-2}{\doi{10.1140/epjc/s10052-012-1896-2}},
\href{http://www.arXiv.org/abs/1111.6097}{\texttt{arXiv:1111.6097}}.

\bibitem{Khachatryan:2016kdb}
\hrefCMSnoop {}{{CMS Collaboration}, ``Jet energy scale and resolution in the
  {CMS} experiment in {$\Pp\Pp$} collisions at 8 {TeV}'',} \textit{ JINST}
  \textbf{ 12} (2017) P02014,
  \href{http://dx.doi.org/10.1088/1748-0221/12/02/P02014}{\doi{10.1088/1748-0221/12/02/P02014}},
\href{http://www.arXiv.org/abs/1607.03663}{\texttt{arXiv:1607.03663}}.

\bibitem{Chatrchyan:2009hy}
\hrefCMSnoop {}{{CMS Collaboration}, ``Identification and filtering of
  uncharacteristic noise in the {CMS} hadron calorimeter'',} \textit{ JINST}
  \textbf{ 5} (2010) T03014,
  \href{http://dx.doi.org/10.1088/1748-0221/5/03/T03014}{\doi{10.1088/1748-0221/5/03/T03014}},
\href{http://www.arXiv.org/abs/0911.4881}{\texttt{arXiv:0911.4881}}.

\bibitem{cms}
\hrefCMSnoop {}{{CMS Collaboration}, ``The {CMS} experiment at the {CERN
  LHC}'',} \textit{ JINST} \textbf{ 3} (2008) S08004,
  \href{http://dx.doi.org/10.1088/1748-0221/3/08/S08004}{\doi{10.1088/1748-0221/3/08/S08004}}.

\bibitem{totem1}
\hrefCMSnoop {}{{TOTEM} Collaboration, ``The {TOTEM} experiment at the {CERN}
  {Large Hadron Collider}'',} \textit{ JINST} \textbf{ 3} (2008) S08007,
  \href{http://dx.doi.org/10.1088/1748-0221/3/08/S08007}{\doi{10.1088/1748-0221/3/08/S08007}}.

\bibitem{totem2}
\hrefCMSnoop {}{{TOTEM} Collaboration, ``Performance of the {TOTEM} detectors
  at the {LHC}'',} \textit{ Int. J. Mod. Phys. A} \textbf{ 28} (2013) 1330046,
  \href{http://dx.doi.org/10.1142/S0217751X13300469}{\doi{10.1142/S0217751X13300469}},
  \href{http://www.arXiv.org/abs/1310.2908}{\texttt{arXiv:1310.2908}}.

\bibitem{lhc_optics}
\hrefCMSnoop {}{{TOTEM} Collaboration, ``{LHC} optics measurement with proton
  tracks detected by the roman pots of the {TOTEM} experiment'',} \textit{ New
  J. Phys.} \textbf{ 16} (2014) 103041,
  \href{http://dx.doi.org/10.1088/1367-2630/16/10/103041}{\doi{10.1088/1367-2630/16/10/103041}},
  \href{http://www.arXiv.org/abs/1406.0546}{\texttt{arXiv:1406.0546}}.

\bibitem{Niewiadomski:2008zz}
\href {http://weblib.cern.ch/abstract?CERN-THESIS-2008-080}{H.~Niewiadomski,
  ``Reconstruction of protons in the {TOTEM} roman pot detectors at the
  {LHC}''}.
\newblock PhD thesis, University of Manchester, 2008.
\newblock
\href{http://cds.cern.ch/record/1131825}{\texttt{cds.cern.ch/record/1131825}}.

\bibitem{totem3}
\hrefCMSnoop {}{{TOTEM} Collaboration, ``Evidence for non-exponential elastic
  proton-proton differential cross-section at low $|t|$ and $\sqrt{s} = 8$
  {TeV} by {TOTEM}'',} \textit{ Nucl. Phys. B} \textbf{ 899} (2015) 527,
  \href{http://dx.doi.org/10.1016/j.nuclphysb.2015.08.010}{\doi{10.1016/j.nuclphysb.2015.08.010}},
  \href{http://www.arXiv.org/abs/1503.08111}{\texttt{arXiv:1503.08111}}.

\bibitem{pythia6}
\hrefCMSnoop {}{T.~Sj{\"o}strand, S.~Mrenna, and P.~Skands, ``{PYTHIA} 6.4
  physics and manual'',} \textit{ JHEP} \textbf{ 05} (2006) 026,
  \href{http://dx.doi.org/10.1088/1126-6708/2006/05/026}{\doi{10.1088/1126-6708/2006/05/026}},
\href{http://www.arXiv.org/abs/hep-ph/0603175}{\texttt{arXiv:hep-ph/0603175}}.

\bibitem{pythia8}
\hrefCMSnoop {}{T.~Sj{\"o}strand, S.~Mrenna, and P.~Skands, ``A brief
  introduction to {PYTHIA} 8.1'',} \textit{ Comput. Phys. Commun.} \textbf{
  178} (2008) 852,
  \href{http://dx.doi.org/10.1016/j.cpc.2008.01.036}{\doi{10.1016/j.cpc.2008.01.036}},
\href{http://www.arXiv.org/abs/0710.3820}{\texttt{arXiv:0710.3820}}.

\bibitem{herwig6}
G.~Corcella\hrefCMSnoop {}{ {et~al.}, ``{HERWIG 6}: An event generator for
  hadron emission reactions with interfering gluons (including supersymmetric
  processes)'',} \textit{ JHEP} \textbf{ 01} (2001) 010,
  \href{http://dx.doi.org/10.1088/1126-6708/2001/01/010}{\doi{10.1088/1126-6708/2001/01/010}},
\href{http://www.arXiv.org/abs/hep-ph/0011363}{\texttt{arXiv:hep-ph/0011363}}.

\bibitem{Field:2010bc}
\href
  {https://inspirehep.net/record/873443/files/arXiv:1010.3558.pdf}{R.~Field,
  ``Early {LHC} underlying event data - findings and surprises'',} in \textit{
  {Hadron collider physics. Proceedings, 22nd Conference, HCP 2010, Toronto,
  Canada, August 23-27, 2010}}.
\newblock 2010.
\newblock
\href{http://www.arXiv.org/abs/1010.3558}{\texttt{arXiv:1010.3558}}.
\newblock

\bibitem{Corke:2010yf}
\hrefCMSnoop {}{R.~Corke and T.~Sj{\"o}strand, ``Interleaved parton showers and
  tuning prospects'',} \textit{ JHEP} \textbf{ 03} (2011) 032,
  \href{http://dx.doi.org/10.1007/JHEP03(2011)032}{\doi{10.1007/JHEP03(2011)032}},
\href{http://www.arXiv.org/abs/1011.1759}{\texttt{arXiv:1011.1759}}.

\bibitem{Khachatryan:2015pea}
\hrefCMSnoop {}{{CMS Collaboration}, ``Event generator tunes obtained from
  underlying event and multiparton scattering measurements'',} \textit{ Eur.
  Phys. J. C} \textbf{ 76} (2016) 155,
  \href{http://dx.doi.org/10.1140/epjc/s10052-016-3988-x}{\doi{10.1140/epjc/s10052-016-3988-x}},
\href{http://www.arXiv.org/abs/1512.00815}{\texttt{arXiv:1512.00815}}.

\bibitem{pomwig}
\hrefCMSnoop {}{{B. Cox and J. Forshaw}, ``{Pomwig}: {Herwig} for diffractive
  interactions'',} \textit{ Comput. Phys. Commun.} \textbf{ 144} (2002) 104,
  \href{http://dx.doi.org/10.1016/S0010-4655(01)00467-2}{\doi{10.1016/S0010-4655(01)00467-2}},
  \href{http://www.arXiv.org/abs/hep-ph/0010303}{\texttt{arXiv:hep-ph/0010303}}.

\bibitem{Navin:2010kk}
\hrefCMSnoop {}{S.~Navin, ``{Diffraction in PYTHIA}'',} (2010).
\href{http://www.arXiv.org/abs/1005.3894}{\texttt{arXiv:1005.3894}}.

\bibitem{Rasmussen:2015qgr}
\hrefCMSnoop {}{C.~O. Rasmussen and T.~Sj{\"o}strand, ``Hard diffraction with
  dynamic gap survival'',} \textit{ JHEP} \textbf{ 02} (2016) 142,
  \href{http://dx.doi.org/10.1007/JHEP02(2016)142}{\doi{10.1007/JHEP02(2016)142}},
\href{http://www.arXiv.org/abs/1512.05525}{\texttt{arXiv:1512.05525}}.

\bibitem{Chatrchyan:2014qka}
\hrefCMSnoop {}{{CMS and TOTEM} Collaboration, ``{Measurement of pseudorapidity
  distributions of charged particles in proton-proton collisions at $\sqrt{s}$
  = 8 TeV by the CMS and TOTEM experiments}'',} \textit{ Eur. Phys. J. C}
  \textbf{ 74} (2014) 3053,
  \href{http://dx.doi.org/10.1140/epjc/s10052-014-3053-6}{\doi{10.1140/epjc/s10052-014-3053-6}},
\href{http://www.arXiv.org/abs/1405.0722}{\texttt{arXiv:1405.0722}}.

\bibitem{Sirunyan:2018zdc}
\hrefCMSnoop {}{{CMS Collaboration}, ``{Measurement of charged particle spectra
  in minimum-bias events from proton-proton collisions at
  $\sqrt{s}=13\,\text{TeV}$}'',} \textit{ Eur. Phys. J. C} \textbf{ 78} (2018)
  697,
  \href{http://dx.doi.org/10.1140/epjc/s10052-018-6144-y}{\doi{10.1140/epjc/s10052-018-6144-y}},
\href{http://www.arXiv.org/abs/1806.11245}{\texttt{arXiv:1806.11245}}.

\bibitem{Agostinelli:2002hh}
\hrefCMSnoop {}{{GEANT4} Collaboration, ``{\GEANTfour}---a simulation
  toolkit'',} \textit{ Nucl. Instrum. Meth. A} \textbf{ 506} (2003) 250,
\href{http://dx.doi.org/10.1016/S0168-9002(03)01368-8}{\doi{10.1016/S0168-9002(03)01368-8}}.

\bibitem{Khachatryan:2016bia}
\hrefCMSnoop {}{{CMS Collaboration}, ``The {CMS} trigger system'',} \textit{
  JINST} \textbf{ 12} (2017) P01020,
  \href{http://dx.doi.org/10.1088/1748-0221/12/01/P01020}{\doi{10.1088/1748-0221/12/01/P01020}},
\href{http://www.arXiv.org/abs/1609.02366}{\texttt{arXiv:1609.02366}}.

\bibitem{bayes}
\hrefCMSnoop {}{{G. D'Agostini}, ``A multidimensional unfolding method based on
  {Bayes'} theorem'',} \textit{ Nucl. Instrum. Meth. A} \textbf{ 362} (1995)
  487,
  \href{http://dx.doi.org/10.1016/0168-9002(95)00274-X}{\doi{10.1016/0168-9002(95)00274-X}}.

\bibitem{Chatrchyan:2011wm}
\hrefCMSnoop {}{{CMS Collaboration}, ``Measurement of energy flow at large
  pseudorapidities in {$\Pp\Pp$} collisions at $\sqrt{s} = 0.9$ and 7 {TeV}'',}
  \textit{ JHEP} \textbf{ 11} (2011) 148,
  \href{http://dx.doi.org/10.1007/JHEP11(2011)148}{\doi{10.1007/JHEP11(2011)148}},
  \href{http://www.arXiv.org/abs/1110.0211}{\texttt{arXiv:1110.0211}}.
[Erratum: \DOI{10.1007/JHEP02(2012)055}].

\end{thebibliography}\endgroup

\cleardoublepage \appendix\section{The CMS Collaboration \label{app:collab}}\begin{sloppypar}\hyphenpenalty=5000\widowpenalty=500\clubpenalty=5000\vskip\cmsinstskip
\textbf{Yerevan Physics Institute, Yerevan, Armenia}\\*[0pt]
A.M.~Sirunyan, A.~Tumasyan
\vskip\cmsinstskip
\textbf{Institut f\"{u}r Hochenergiephysik, Wien, Austria}\\*[0pt]
W.~Adam, F.~Ambrogi, E.~Asilar, T.~Bergauer, J.~Brandstetter, M.~Dragicevic, J.~Er\"{o}, A.~Escalante~Del~Valle, M.~Flechl, R.~Fr\"{u}hwirth\cmsAuthorMark{1}, V.M.~Ghete, J.~Hrubec, M.~Jeitler\cmsAuthorMark{1}, N.~Krammer, I.~Kr\"{a}tschmer, D.~Liko, T.~Madlener, I.~Mikulec, N.~Rad, H.~Rohringer, J.~Schieck\cmsAuthorMark{1}, R.~Sch\"{o}fbeck, M.~Spanring, D.~Spitzbart, W.~Waltenberger, J.~Wittmann, C.-E.~Wulz\cmsAuthorMark{1}, M.~Zarucki
% ...
\vskip\cmsinstskip
% ...
1:  Also at Vienna University of Technology, Vienna, Austria\\
% ...
\section{The TOTEM Collaboration\label{app:totem}}
\newcommand{\AddAuthor}[2]{#1$^{#2}$,\ }
\newcommand\AddAuthorLast[2]{#1$^{#2}$}
\noindent
    \AddAuthor{G.~Antchev}{a}%
	\AddAuthor{P.~Aspell}{9}%
	\AddAuthor{I.~Atanassov}{a}%
	\AddAuthor{V.~Avati}{7,9}%
	\AddAuthor{J.~Baechler}{9}%
	\AddAuthor{C.~Baldenegro~Barrera}{11}%
	\AddAuthor{V.~Berardi}{4a,4b}%
	\AddAuthor{M.~Berretti}{2a}%
    \AddAuthor{V.~Borchsh}{8}%
	\AddAuthor{E.~Bossini}{9,6b}%
	\AddAuthor{U.~Bottigli}{6b}%
	\AddAuthor{M.~Bozzo}{5a,5b}%
    \AddAuthor{H.~Burkhardt}{9}%
	\AddAuthor{F.~S.~Cafagna}{4a}%
	\AddAuthor{M.~G.~Catanesi}{4a}%
	\AddAuthor{M.~Csan\'{a}d}{3a,b}%
	\AddAuthor{T.~Cs\"{o}rg\H{o}}{3a,3b}%
	\AddAuthor{M.~Deile}{9}%
	\AddAuthor{F.~De~Leonardis}{4c,4a}%
	\AddAuthor{M.~Doubek}{1c}%
	\AddAuthor{D.~Druzhkin}{8,9}%
	\AddAuthor{K.~Eggert}{10}%
	\AddAuthor{V.~Eremin}{d}%
	\AddAuthor{A.~Fiergolski}{9}%
	\AddAuthor{L.~Forthomme}{2a,2b}%
	\AddAuthor{F.~Garcia}{2a}%
	\AddAuthor{V.~Georgiev}{1a}%
	\AddAuthor{S.~Giani}{9}%
	\AddAuthor{L.~Grzanka}{7}%
	\AddAuthor{J.~Hammerbauer}{1a}%
	\AddAuthor{T.~Isidori}{11}%
	\AddAuthor{V.~Ivanchenko}{8}%
	\AddAuthor{M.~Janda}{1c}%
	\AddAuthor{A.~Karev}{9}%
	\AddAuthor{J.~Ka\v{s}par}{1b,9}%
    \AddAuthor{B.~Kaynak}{e}%
	\AddAuthor{J.~Kopal}{9}%
	\AddAuthor{V.~Kundr\'{a}t}{1b}%
	\AddAuthor{S.~Lami}{6a}%
	\AddAuthor{R.~Linhart}{1a}%
	\AddAuthor{C.~Lindsey}{11}%
	\AddAuthor{M.~V.~Lokaj\'{\i}\v{c}ek}{1b,\dagger}%
	\AddAuthor{L.~Losurdo}{6b}%
	\AddAuthor{F.~Lucas~Rodr\'{i}guez}{9}%
	\AddAuthor{M.~Macr\'{\i}}{5a}%
	\AddAuthor{M.~Malawski}{7}%
	\AddAuthor{N.~Minafra}{11}%
	\AddAuthor{S.~Minutoli}{5a}%
	\AddAuthor{T.~Naaranoja}{2a,2b}%
	\AddAuthor{F.~Nemes}{9,3a}%
	\AddAuthor{H.~Niewiadomski}{10}%
	\AddAuthor{T.~Nov\'{a}k}{3b}%
	\AddAuthor{E.~Oliveri}{9}%
	\AddAuthor{F.~Oljemark}{2a,2b}%
	\AddAuthor{M.~Oriunno}{f}%
	\AddAuthor{K.~\"{O}sterberg}{2a,2b}%
	\AddAuthor{P.~Palazzi}{9}%
	\AddAuthor{V.~Passaro}{4c,4a}%
	\AddAuthor{Z.~Peroutka}{1a}%
	\AddAuthor{J.~Proch\'{a}zka}{1b}%
	\AddAuthor{M.~Quinto}{4a,4b}%
	\AddAuthor{E.~Radermacher}{9}%
	\AddAuthor{E.~Radicioni}{4a}%
	\AddAuthor{F.~Ravotti}{9}%
	\AddAuthor{C.~Royon}{11}%
	\AddAuthor{G.~Ruggiero}{9}%
	\AddAuthor{H.~Saarikko}{2a,2b}%
    \AddAuthor{V.D.~Samoylenko}{c}%
	\AddAuthor{A.~Scribano}{6a}%
	\AddAuthor{J.~Siroky}{1a}%
	\AddAuthor{J.~Smajek}{9}%
	\AddAuthor{W.~Snoeys}{9}%
	\AddAuthor{R.~Stefanovitch}{9}%
	\AddAuthor{J.~Sziklai}{3a}%
	\AddAuthor{C.~Taylor}{10}%
	\AddAuthor{E.~Tcherniaev}{8}%
	\AddAuthor{N.~Turini}{6b}%
    \AddAuthor{O.~Urban}{1a}%
	\AddAuthor{V.~Vacek}{1c}%
    \AddAuthor{O.~Vavroch}{1a}%
	\AddAuthor{J.~Welti}{2a,2b}%
	\AddAuthor{J.~Williams}{11}%
	\AddAuthor{J.~Zich}{1a}%
	\AddAuthorLast{K.~Zielinski}{7}%

\vskip 4pt plus 4pt
\let\thefootnote\relax
\newcommand{\AddInstitute}[2]{${}^{#1}$#2\\}
\newcommand{\AddExternalInstitute}[2]{\footnote{${}^{#1}$ #2}}
\noindent
	\AddInstitute{1a}{University of West Bohemia, Pilsen, Czech Republic.}
	\AddInstitute{1b}{Institute of Physics of the Academy of Sciences of the Czech Republic, Prague, Czech Republic.}
	\AddInstitute{1c}{Czech Technical University, Prague, Czech Republic.}
	\AddInstitute{2a}{Helsinki Institute of Physics, University of Helsinki, Helsinki, Finland.}
	\AddInstitute{2b}{Department of Physics, University of Helsinki, Helsinki, Finland.}
	\AddInstitute{3a}{Wigner Research Centre for Physics, RMKI, Budapest, Hungary.}
	\AddInstitute{3b}{EKU KRC, Gy\"ongy\"os, Hungary.}
	\AddInstitute{4a}{INFN Sezione di Bari, Bari, Italy.}
	\AddInstitute{4b}{Dipartimento Interateneo di Fisica di Bari, Bari, Italy.}
	\AddInstitute{4c}{Dipartimento di Ingegneria Elettrica e dell'Informazione --- Politecnico di Bari, Bari, Italy.}
	\AddInstitute{5a}{INFN Sezione di Genova, Genova, Italy.}
	\AddInstitute{5b}{Universit\`{a} degli Studi di Genova, Italy.}
	\AddInstitute{6a}{INFN Sezione di Pisa, Pisa, Italy.}
	\AddInstitute{6b}{Universit\`{a} degli Studi di Siena and Gruppo Collegato INFN di Siena, Siena, Italy.}
	\AddInstitute{7}{AGH University of Science and Technology, Krakow, Poland.}
	\AddInstitute{8}{Tomsk State University, Tomsk, Russia.}
	\AddInstitute{9}{CERN, Geneva, Switzerland.}
	\AddInstitute{10}{Case Western Reserve University, Dept.~of Physics, Cleveland, OH, USA.}
	\AddInstitute{11}{The University of Kansas, Lawrence, USA.}

	\AddExternalInstitute{a}{INRNE-BAS, Institute for Nuclear Research and Nuclear Energy, Bulgarian Academy of Sciences, Sofia, Bulgaria.}
	\AddExternalInstitute{b}{Department of Atomic Physics, ELTE University, Budapest, Hungary.}
    \AddExternalInstitute{c}{NRC `Kurchatov Institute'--IHEP, Protvino, Russia.}
	\AddExternalInstitute{d}{Ioffe Physical - Technical Institute of Russian Academy of Sciences, St.~Petersburg, Russian Federation.}
	\AddExternalInstitute{e}{Istanbul University, Istanbul, Turkey.}
	\AddExternalInstitute{f}{SLAC National Accelerator Laboratory, Stanford CA, USA.}
	\AddExternalInstitute{\dagger}{Deceased.}
\addtocounter{footnote}{-8}
\newcommand{\thefootnote}{\alph{footnote}} \end{sloppypar}
\end{document}